\documentclass[aps,showpacs,amsmath,amssymb,nofootinbib,dvips,showkeys,preprintnumbers]{revtex4}

\usepackage{graphicx}   
\usepackage{color}      
\usepackage{braket}
\usepackage{tikz}
\usepackage{pgfplots}

\def \be  {\begin{equation}}
\def \ee  {\end{equation}}
\def \ee  {\end{equation}}
\def \bea {\begin{eqnarray}}
\def \eea {\end{eqnarray}}

\begin{document}
\preprint{ECTP-2019-07}    
\preprint{WLCAPP-2019-07}
\hspace{0.05cm}

\title{Bulk viscosity in strong and electroweak matter}

\author{Abdel Nasser Tawfik} 
\email{tawfik@itp.uni-frankfurt.de}
\affiliation{Goethe University, Institute for Theoretical Physics (ITP), Max-von-Laue-Str. 1, D-60438 Frankfurt am Main, Germany}
\affiliation{Egyptian Center for Theoretical Physics (ECTP), Juhayna Square of 26th-July-Corridor, 12588 Giza, Egypt}

\author{Carsten Greiner} 
\affiliation{Goethe University, Institute for Theoretical Physics (ITP), Max-von-Laue-Str. 1, D-60438 Frankfurt am Main, Germany}

\date{\today}

\begin{abstract}

For temperatures $T$ ranging from a few MeV up to TeV and energy density $\rho$ up to $10^{16}~$GeV/fm$^3$, the bulk viscosity $\zeta$ is calculated in non--perturbation (up, down, strange, charm, and bottom) and perturbation theories with up, down, strange, charm, bottom, and top quark flavors, at vanishing baryon--chemical potential. To these calculations, results deduced from the effective QCD--like model, the Polyakov linear--sigma model (PLSM), are also integrated in. The PLSM merely comes up with essential contributions for the vacuum and thermal condensations of the gluons and the quarks (up, down, strange, and charm flavors). Furthermore, the thermal contributions of the photons, neutrinos, charged leptons, electroweak particles, and scalar Higgs boson, are found very significant along the entire range of $T$ and $\rho$ and therefore could be well integrated in. We present the dimensionless quantity $9 \omega_0 \zeta/Ts$, where $\omega_0$ is a perturbative scale and $s$ is the entropy density and conclude that $9 \omega_0 \zeta/Ts$ exponentially decreases with increasing $T$. We also conclude that the resulting $\zeta$ with the non--perturbative and perturbative QCD contributions non--monotonically increases with increasing $\rho$. But with nearly--entire standard model contributions considered in the present study, $\zeta$ almost--linearly increases with increasing $\rho$. Apparently, these results offer a great deal to explore in astrophysics, cosmology, and nuclear collisions.

\end{abstract}


\maketitle

\tableofcontents
\makeatletter
\let\toc@pre\relax
\let\toc@post\relax
\makeatother 


\section{Introduction \label{intro}}

Various high--energy experiments, for instance, at the Super Proton Synchrotron (SPS) at CERN, at the Relativistic Heavy Ion Collider (RHIC) at BNL and at the Large Hadron Collider (LHC) at CERN have collected unambiguous evidences for the strongly--correlated viscous quantum choromodynamic (QCD) matter; the quark--gluon plasma (QGP), the colorless state of partonic matter \cite{SHURYAK198071,Gyulassy:2004zy,Heinz:2011kt,Adamczyk:2013dal,Ryu:2017qzn,Bzdak:2019pkr}. The discovery of the new state of QCD matter, the QGP, in 1999 could be credited to SPS, at CERN \cite{Tawfik:2000mw,Heinz:2000ba}. The RHIC discovery announced in 2004 was about a highly viscous property of the QGP \cite{Gyulassy:2004zy}. The LHC did confirm both SPS and RHIC discovery \cite{Heinz:2011kt,Ryu:2017qzn}, which was also strenthened by the beam energy scan (BES) programm at RHIC \cite{Adamczyk:2013dal,Bzdak:2019pkr}. 

For perturbative gauge QCD, it was suggested that the shear viscosity normalized to the entropy density is very close to the lower bound predicted by Anti--de Sitter/Conformal Field Theory (AdS/CFT) \cite{Kovtun:2004de}. The non--perturbative estimations for the viscous properties date back to 2007 and have been found to highlight an unambiguous temperature dependence \cite{Sakai:2007cm,PhysRevD.98.054515}. In all these studies, the range of temperature is limited to a few times the QCD scale or the critical temperature. The present manuscript introduces bulk viscosity at temperatures ($T$) ranging from a few MeV up to TeV or at energy density ($\rho$) up to $10^{16}~$GeV/fm$^3$ including all known elementary particles from quarks to Higgs bosons, for which astrophysical and cosmological implications are very likely. The direct implication on nuclear collisions is also obvious.

From the {\it discrete} Green function in variables of the Matsubara frequencies, the first--principle simulations for the shear and bulk viscosity have been calculated on isotropic $24^3 \times 8$ and $16^3 \times 8$ lattices \cite{Sakai:2007cm} A new lattice study on the temperature dependence of the bulk viscosity of SU($3$) gluodynamics on $48^3 \times 16$ lattice was reported, recently \cite{PhysRevD.98.054515}. The viscous coefficients are then determined from the slopes of the spectral functions, at vanishing Matsubara frequency. From the retarded Green function defined by the Kramers--Kronkig relation and given in terms of the thermodynamic quantities \cite{Karsch:2007jc}, the bulk viscosity could be interpretted as a measure for the violation of the conformal invariance. Because QCD at the classical level is conformally invariant, the quark and gluon condensates have been assumed to significantly contribute to the bulk viscosity \cite{Tawfik:2016ihn,Tawfik:2016edq}. This is the reason why we are introducing vacuum and thermal contributions from the quark (u--, d--, s--, and c--quarks) and gluon condensates \cite{Miller:2003ha,Miller:2003hh,Miller:2003ch,Miller:2004uc} to the bulk viscosity, besides the contributions from non--perturbative (u--, d--, s--, c-- and b--quark) and perturbative (u--, d--, s--, c--, b--, and t--quark) theories \cite{PhysRevD.98.054515,Borsanyi:2016ksw,Laine:2015kra,DOnofrio:2015gop}. For the sake of completeness, we emphasize that these are not the only contributions we take into contributions. The thermal contributions of the photons, neutrinos, charged leptons, electroweak particles, and the scalar Higgs boson, are found even very significant, along the entire range of temperatures and energy densities. Accordingly, we believe that the present study takes into account the up--to--date maximal degrees--of--freedom of the standard model for elementary particles.

The quark condensates in non--perturbative QCD--like models with as much as possible quark flavors (u--, d--, s--, and c--quarks) have been estimated \cite{Roder:2003uz,AbdelAalDiab:2018hrx}. On the other hand, various thermodynamic quantities have been calculated in most reliable non--perturbative with $2+1+1+1$ ($3+1+1$) and perturbative simulations with $2+1+1+1+1$ ($3+1+1+1$) quark flavors, at vanishing baryon density \cite{Borsanyi:2016ksw,Laine:2015kra,DOnofrio:2015gop,Tawfik:2019jsa}. Taking into account the contributions of gauge bosons: photons, $W^{\pm}$, and $Z^0$, charged leptons: neutrino, electron, muon, and tau, and the Higgs bosons: scalar Higgs particle, we introduce the temperature and energy--density dependence of the bulk viscosity for almost the whole constituents of the standard model for elementary particles. Again, that the temperatures range from a few MeV to TeV or the energy density from a few hundreds MeV/fm$^3$ to $10^{16}~$GeV/fm$^3$, makes implications of the present studies on astrophysics, cosmology, and nuclear collisions, for instance, very likely. Such a wide range (of $T$ and $\rho$) does not only cover both strong and electroweak (EW) phases, which obviously become nowadays - to a large extend - accessible by high--energy experiments \cite{Hu:2017pat} and recent astrophysical observations \cite{Planck:2018vyg}, but also goes beyond that and approaches domains, which are likely never accessed, so far. For examples, the additional contributions by the gauge bosons, the charged leptons, and the Higgs bosons make it possible for the energy density to approach two--order--of--magnetiude GeV/fm$^3$ larger than the energy density reached in perturpation theory \cite{Laine:2015kra}. Due to this fact, the present manuscript together with ref. \cite{Tawfik:2019jsa} complete the set of barotropic equations--of--state needed for the early Universe, for instance, enabling us to solve Friedman equations for viscous cosmic background. Actually, this motivates the current study.

Another aspect of applying finite bulk viscosity is the cosmological inflation in the Big Bang Theory. The bulk viscosity as calculated during the GUT phase transition was suggested to contribute to the cosmological inflation and density fluctuations \cite{Cheng:1991uu}. The conditions under which the bulk viscosity becomes an inflationary source have been evaluated, as well. It was assumed that the finite bulk viscosity considerably contributes to resolving the entropy generation problem \cite{Cheng:1991uu,Tawfik:2009mk}. The evolution of the early Universe as described by a spatially homogeneous and isotropic Robertson--Walker model strongly depends on whether or not the viscous coefficients (bulk and/or shear) are taken finite \cite{Singh:2008zzj,Tawfik:2011sh,Tawfik:2010bm,Tawfik:2010pm}. 

For nuclear collisions, the other implication of the present study, a possible mechanism for the simultaneous radial flow and azimuthal anisotropy in high--nergy collisions \cite{Adamczyk:2017ird} was suggested as finite bulk viscosity \cite{Ryu:2017qzn}. Also, the interferometry correlations, i.e., the shapes of the freeze--out hyper--surface or the outer layers of the fireball and the related reduction of the ratio of two interferometry radii, can only be explained by temperature--depending bulk viscosity \cite{Bozek:2017kxo}. These are few examples on the roles that the bulk viscosity would play in nuclear collisions during either hadronic \cite{Ryu:2017qzn,Adamczyk:2017iwn,Bozek:2017kxo,Tawfik:2010mb,Chakraborty:2010fr} or partonic (QGP) phases \cite{Sakai:2007cm,Karsch:2007jc}. 

The present script is organized as follows. The various approaches for calculating the bulk viscosity are introduced in section \ref{sec:apprs}. Both Boltzmann--Uehling--Uhlenbeck (BUU) and the Green--Kubo (GK) equations are reviewed in section \ref{sec:buu} and \ref{sec:GK}, respectively. The quark and the gluon condensates are outlined in section \ref{sec:qrkCondns1} and \ref{sec:glnCondns1}, respectively. The first--principle approach, namely the quantum chromodynamics (QCD), is discussed in section \ref{sec:qcd}, where the QCD--like model, the Polyakov linear--sigma model (PLSM) is also presented in section \ref{sec:PLSMbv} and the lattice QCD simulations are given in section \ref{sec:LQCDbv}. The results and the discussion are elaborated in section \ref{resulat}. We start with the bulk viscosity in PLSM, section \ref{sec:hrmplsm}, and then present the bulk viscosity in non--perturbative and perturbative calculations in section \ref{sec:bvlqcd}. The latter is divided into QCD contributions, section \ref{sec:QCDconts}, and the SM contributions, \ref{sec:SMconts}. Section \ref{conclusion} is devoted to the final conclusions.

\section{Theoretical approach}
\label{sec:apprs}

For a system characterized by quark degrees of freedom, the equilibrium energy--momentum tensor can be given as \cite{Weinberg:1995mt} 
\bea
T^{\mu\nu} &=& -p\, g^{\mu \nu} + \mathcal{H}\, u^\mu\, u^\nu + \Delta  T^{\mu\nu}, \label{Tensor0}
\eea
where $u^{\nu|\mu}$ are four velocities and $p$ stands the thermodynamic pressure. The enthalpy density $\mathcal{H}=p + \rho$ can also be expressed in dependence on the energy density $\rho = -p + Ts$ and the entropy $s$. When inserting a dissipative part into Eq. (\ref{Tensor0}), this is then derived towards out--of--equilibrium 
\bea
\Delta \, T^{\mu\nu} &=& \eta \Big( D^\mu u^\nu +  D^\nu u^\mu + \frac{2}{3} \Delta^{\mu\nu} \partial_\sigma u^\sigma \Big) - \zeta \Delta^{\mu\nu} \partial_\sigma u^\sigma. \label{disspative}
\eea
In this expression, the Landau--Lifshitz condition is fulfilled, $u_\mu \, \Delta  T^{\mu\nu}=0$ \cite{Weinberg:1995mt} and in local rest--frame, the hydrodynamic expansion reads \cite{Weinberg:1995mt}
\bea
\delta T^{ij } &=&\sum_f \int d \Gamma^* \frac{p^i\, p^j}{E_f}  \Big[ -\mathcal{A}_f \,\partial_\sigma u^\sigma - \mathcal{B}_f\, p_f^\nu D_\nu \left( \frac{\mu}{T}\right) 
+ \mathcal{C}_f\, p_f^\mu  p_f^\nu   \Big( D^\mu u^\nu +  D^\nu u^\mu + \frac{2}{3} \Delta^{\mu\nu} \partial_\sigma u^\sigma \Big) \Big] f_f^{eq}. \label{delTij5}
\eea
The sum runs over the degrees of freedom, such as the quarks and the antiquarks. $\mathcal{A}_f,\, \mathcal{B}_f$ and $\mathcal{C}_f$ are functions depending on the momentum $p$. $d \Gamma^*$ representing a generic phase-space.  It is obvious that in the local rest--frame, the given derivative vanishes, i.e., $\partial_{k} u_0=0$. This means that the sum over $\mu$ and $\nu$ is equivalent to the sum over the spatial indices $\rho$ and $\sigma$. This leads to $p_f^i  p_f^j  p_f^\sigma  p_f^\rho=|p_f|^4 (\delta_{ij} \delta_{\sigma \rho}+\delta_{i\sigma} \delta_{j \rho} +\delta_{i\rho} \delta_{j\sigma}$). Also, in the local rest--frame, $p_f=p$. When assuming that both Eqs. (\ref{disspative}) and (\ref{Tensor0}) are equal, then the dissipation parts of the energy--momentum tensor can be determined, straightforwardly. For all these reasons, it seems of a great advantageous to assume a local rest--frame of the fluid of interest. 

The transport properties are defined as the coefficients of the spatial components of the difference between the equilibrium and out--of--equilibrium energy--momentum tensors with respect to the Lagrangian density \cite{Tawfik:2010mb}. In section \ref{sec:blkvsc1}, we discuss well--known approaches for the viscous coefficients, namely the Boltzmann--Uehling-Uhlenbeck (BUU), section \ref{sec:buu} and the Green--Kubo (GB) approach, section \ref{sec:GK}.

\subsection{Formula for bulk viscosity}
 \label{sec:blkvsc1}
 
\subsubsection{Boltzmann-Uehling-Uhlenbeck (BUU) Approach}
\label{sec:buu}

For an equilibrium state having $f$ quark flavors with momenta $\vec{p}$, the phase--space distribution for these fermions including the Polyakov loop variables can be expressed as  \cite{Tawfik:2016ihn} \footnote{It should be noticed that ref. \cite{Tawfik:2016ihn} introduces an expression for the Polyakov potential which differs from the one utilized in the present study. Such a slight difference doesn't significantly affect both results and  Polyakov loops $\phi$ and $\phi^{\ast}$ \cite{Tawfik:2016edq}.}
\begin{eqnarray}
f_f (T,\mu) &=&\ln \left[ 1+3\left(\phi+\phi^{\ast} \,e^{-\frac{E_f-\mu _f}{T}}\right)\times e^{-\frac{E_f-\mu _f}{T}}+e^{-3 \frac{E_f-\mu _f}{T}}\right]. \label{fqaurk}  
\end{eqnarray}
where $E_f =(m_f^2 +p^2)^{1/2}$ is the dispersion relation of $f$--th quark flavor. A similar expression can be deduced for anti--quarks, where $\phi$ and $\phi^{\ast}$ are exchanged and $-\mu_f\rightarrow +\mu_f$. We also express Eq. (\ref{fqaurk}) in terms of $\phi$ and $\phi^{\ast}$, the Polyakov loop variables, Eq. (\ref{phis}). 
As $\phi$ and $\phi^{\ast}\rightarrow 0$, Eq. (\ref{fqaurk}) merely loses the Polyakov contributions and apparently retains its standard expression for the phase-space distribution of the quarks, the fermions. This expression shall be needed when utilizing PLSM, section \ref{sec:PLSMbv}, in order to determine the bulk viscosity and the temperature dependence of the different quark condensates, section \ref{sec:qrkCondns1}.

At finite temperature and density, the relaxation time approximation, for example, can be applied to the Boltzmann--Uehling--Uhlenbeck (BUU) expression \cite{Chakraborty:2010fr} with Chapman--Enskog expansion. In non--Abelian external field, the viscous coefficients can be estimated from relativistic kinetic theory. As we focus on the bulk viscosity, this is given as \cite{Chakraborty:2010fr},
\bea
\zeta (T,\mu) &=&  \frac{1}{9T}  \sum_{f} \int \frac{d^3p}{(2\pi)^3} \, \frac{\tau _f}{E_f ^2}\, \left[\frac{|\vec{p}|^2}{3} - c_s^2 E_f^2 \right]^2 \, f_f (T, \mu). \label{eq:zetaBUU1}
\eea
For macroscopic consideration, the relaxation time $\tau _f$, which involves complicated collision integrals, could be - for simplicity - determined as the mean collision time and thus from thermal averages \cite{Tawfik:2016edq,Tawfik:2011sh,Tawfik:2010bm}
\bea
\tau_f(T) &=& \frac{1}{n_f(T)\, \langle v(T)\rangle \sigma(T)}, \label{eq:tau1}
\eea
where $\langle v(T)\rangle$ is the mean relative velocity of two colliding particles, $\sigma$ is the cross section, and $n_f(T)$ is the corresponding number density. Approaches to determine $\langle v(T)\rangle$ and $\sigma$ habe been discussed in ref. \cite{Tawfik:2010bm}, for instance.

When adding a small perturbation, i.e., even the local equilibrium is slightly derived towards an out--of--equilibrium status, the four velocity $u^{\mu} (x)$ becomes no longer constant in space and time, then the energy--momentum tensor $T^{\mu \nu}$ and the distribution function $f_f (T,\mu)$, for instance, depart from the thermal equilibrium, as well,
\bea
\delta T^{j j} &=&  \sum_f \int \frac{d^3p}{(2\pi)^3} \frac{p^i \,p^j}{E_f} f_f ^{eq} (u_i\, p^i/T) \, \phi_f (x,p),  \label{eq:modT1}\\
f_f (x,p) &=& f^{eq} \left(u_i\, p^i/T\right) \Big[1+\phi_f (x,p) \Big],  \label{eq:modf1}\\
\phi_f &=& \Big[ -\mathcal{A}_f \,\partial_\sigma u^\sigma - \mathcal{B}_f\, p_f^\nu D_\nu \left( \frac{\mu}{T}\right) + \mathcal{C}_f\, p_f^\mu  p_f^\nu   \Big( D^\mu u^\nu +  D^\nu u^\mu + \frac{2}{3} \Delta^{\mu\nu} \partial_\sigma u^\sigma \Big) \Big]. \label{eq:modphi1}
\eea
In order to determine $\mathcal{A}_f$ and $\mathcal{C}_f$, we recall the Boltzmann master equation \cite{Chakraborty:2010fr}, which can be expressed as
\bea
\frac{\partial f_f (x,t,p)}{\partial t} &=& \left(\frac{\partial}{\partial t} \, + \frac{\partial}{\partial x^i} \frac{\partial x^i}{\partial t}+ \frac{\partial}{\partial p^i} \frac{\partial p^i}{\partial t}  \right) f_f (x,t,p). 
\eea
The right--hand side gives the collision integral. For a collision such as $\lbrace i\rbrace\leftrightarrow \lbrace j\rbrace$, the equilibrium distribution functions are identical, $f_{\lbrace i\rbrace}^{eq} = f_{\lbrace j\rbrace}^{eq}$ \cite{Chakraborty:2010fr} and the collision integral can be given as
\bea
{\bf C}\,[f_f] &=&  \sum_{\lbrace i\rbrace\lbrace j\rbrace;f}  \frac{1}{S} \int  \Big(\frac{dk_z}{2\pi}\Big)_{\lbrace i\rbrace}\Big(\frac{dk_z}{2\pi}\Big)_{\lbrace j\rbrace}  W(\lbrace i\rbrace |\lbrace j\rbrace) F[f_f],
\eea
in which the statistical factor $S$ takes into consideration identical particles. $F[f_f]$ expresses Bose--Einstein or Fermi--Dirac distribution statistics \cite{Chakraborty:2010fr},
\bea
F[f_f] &=& \Pi_{\lbrace i\rbrace} \Pi_{\lbrace j\rbrace} \left\{f_j\left(1+(-1)^{s_i}f_i\right) - f_i\left(1+(-1)^{s_j}f_j\right) \right\}.
\eea
Due to Landau--Lifshitz condition, some constrains can be added to $\phi_f (x,p)$, Eq. (\ref{eq:modphi1}), so that $|\phi_f| \ll 1$ \cite{Chakraborty:2010fr}. Under these assumptions, a particular solution that conserves the Landau--Lifshitz condition, as well, can be proposed as \cite{Chakraborty:2010fr,Tawfik:2016ihn}. 
\bea
\mathcal{A}_f &=& \mathcal{A}_f^{\mbox{par}} - b E_{f},\\
\mathcal{A}_f^{\mbox{par}} &=& \frac{\tau_f}{3T} \, \left[\frac{|\vec{p}|^2}{3} - c_s^2 E_{f}^2 \right], \label{eq:Apart}
\eea 
where $\tau_f$ is the relaxation time, Eq. (\ref{eq:tau1}), which can be - for instance - linked to the decay width as discussed in refs. \cite{Tawfik:2016edq,Marty:2013ita}, and $c_s^2 =\partial p/\partial\rho$ is the speed of sound squared. 
  
In relaxation time approximation \cite{Tawfik:2016ihn}, the phase--space distributions of quarks and antiquarks can be replaced by their equilibrium ones; $f=f^{eq}+\delta f$, where $\delta f$ is allowed to be arbitrary infinitesimal, while the collision integral can be approximated as ${\bf C}\,[f_f]=\delta f/\tau_f$ \cite{Chakraborty:2010fr}. Then, the particular solution $\mathcal{A}_f^{\mbox{par}}$, Eq. (\ref{eq:Apart}), is also valid. Furthermore, we get \cite{Chakraborty:2010fr},
\bea
\mathcal{C}_f^{\mbox{par}} &=& \frac{\tau_f}{2T E_f}.
\eea

\subsubsection{Green-Kubo (GK) Approach}
\label{sec:GK}

In Lehmann spectral representation of two--point correlation functions of the energy--momentum tensor, the Green--Kubo formalisms for the bulk and the shear viscosity can be expressed  as \cite{Kubo:1957mj,zubarev1974nonequilibrium}
\bea
\left(\begin{array}[c]{c}\zeta  \\ \eta\end{array}
\right)=\lim_{\omega\rightarrow 0^+} \lim_{|{\bf p}|\rightarrow 0^+} \frac{1}{\omega}
\left(\begin{array}[c]{c} \frac{1}{2} A_{\zeta} (\omega, |{\bf p}|) \\ \frac{1}{20} A_{\eta} (\omega, |{\bf p}|)  \end{array}\right), \label{eq:matrix_field_A}
\eea
where $A_{\zeta}$  and $ A_{\eta}$ are spectral functions \cite{Kubo:1957mj}
\bea
A_{\zeta} (\omega, |{\bf p}|)&=& \int d^4 x\; e^{ip\cdot x} \langle\left[\mathcal{P}(x), \mathcal{P}(0)\right] \rangle, \label{eq:Azeta} \\ 
A_{\eta} (\omega, |{\bf p}|) &=& \int d^4 x\; e^{ip\cdot x} \langle\left[\pi^{ij}(x), \pi^{ij}(0)\right] \rangle, \label{eq:Aeta} \\
\mathcal{P}(x) &=& -\frac{1}{3} T^{i}_{i} (x) - c_s^2 T^{00} (x), \\
\pi^{ij}(x)  &=& T^{ij} (x) -\frac{1}{3} \delta^{ij} T^{k}_{k} (x),
\eea
and $\langle\left[\cdots \right] \rangle$ stands for an appropriate thermal averaging. 

For simplicity, we limit the discussion to the shear viscosity, whose derivation is similar to that of the bulk viscosity. When expressing the energy--momentum tensor in terms of the Lagrangian density \cite{Tawfik:2016ihn},  we get
\bea
T^{\mu \nu}= -g^{\mu \nu} \mathcal{L} + \frac{\partial \mathcal{L}}{\partial (\partial_\mu \Phi)} \partial^{\nu} \Phi.
\eea
The viscous stress tensor is then determined by the Lagrangian part, for instance, 
\bea
\pi_{\mu\nu} &=& \left( \Delta_{\mu\nu}  \Delta^{\rho\sigma} -\frac{1}{3} \Delta_{\mu\rho}  \Delta^{\nu\sigma} \right) T^{\rho\sigma},
\eea
where $\Delta^{\mu\nu}=g^{\mu\nu}-u^\mu u^\nu$. 

In linear response theory (LRT), we can estimate the impacts of the dissipative forces on the energy--momentum tensor. It is obvious to expect that these forces are small elative to the typical energies of the system of interest, especially in strongly interacting systems \cite{Ghosh:2014yea}, as what we are dealing with in the present paper. The linear response of the microscopic viscous stress--tensor, for instance, $\pi^{\mu\nu}$, to the dissipative forces, makes it possible to relate the correlation functions to the macroscopic viscosity parameter \cite{Lang:2012tt}. With an appropriate thermal averaging of two--point function, $\langle \cdots \rangle$, two point correlator of the viscous stress--tensor can be deduced as 
\bea
\Pi_{ab}(|{\bf p}|) = i \int d^4x\; e^{ip\cdot x} \langle \tau_c \pi_{\mu \nu} (x) \pi^{\mu\nu} (0) \rangle^{ab},
\eea
where $a$, $b\in [1, 2]$ represents the thermal indices of $2 \times 2$--matrix for $\langle \cdots \rangle$  and $\tau_c$ is time ordering with respect to a contour in the complex time plane.

The spectral function can then be written as
\bea
A_{\eta} (\omega, |{\bf p}|) &=& 2 \tanh \left(\frac{\omega/T}{2}\right)\mbox{Im}\; \Pi_{11} (\omega, p). \label{eq:spctlGK}
\eea
The diagonal element can be related to the retarded two--point function of the viscous stress--tensor. There are $11$ components \cite{Ghosh:2014yea}
\bea
\Pi_{11}(|{\bf p}|) &=& i \int d\Gamma^{\ast} \;N(p,k)\; D_{11}(k)\; D_{11}(p-k), \label{stress}
\eea
where $D_{11}(p)$ is the scalar part of the $11$ components of the quark--propagator matrix and $N(p,k)$ includes the numerator part of the propagators. Further details are now in order. These can be summarized as follows.
\begin{itemize}
\item The $11$ components of the scalar part of the thermal propagator can be expressed by using formalism of real--time thermal field theory (RFT)
\bea
D^{11} (k) &=& \frac{-1}{k_0^2-E_{B,f}^2 + i\rho}- 2\pi i\;  E_{B,f}(E_k)\; f_{f}(k)  \delta(k_0^2 - E_{B,f}^2 (k)).
\eea
When replacing the momentum indices $p\rightarrow k$ in Eq. (\ref{fqaurk}), the Fermi--Dirac distribution function and the modified dispersion relation can be expressed in terms of the Polyakov--loop variables, as well. 

\item The remaining parts in Eq. (\ref{stress}) stand for fermions \cite{zubarev1974nonequilibrium,FernandezFraile:2009mi,Ghosh:2014yea}, where 
\bea
N(p,k) &=&  \frac{32}{3} k_0 (k_0+\omega) {\bf k} \cdot ({\bf k} +{\bf p}) - 4 \Big( {\bf k} \cdot ({\bf k} +{\bf p}) + \frac{1}{3} {\bf k}^2 ({\bf k} +{\bf p})^2 \Big). \label{Nterm}
\eea
\end{itemize}

\begin{center}
\begin{figure}[h]
 \begin{tikzpicture}
 \pgfplotsset{every x tick label/.append style={font=\LARGE, yshift=0.5ex}}
  \draw[-latex] [black,thick,line width=1.5pt,domain=-180:-88] plot ({cos(\x)}, {sin(\x)});
 \draw [black,thick,line width=1.5pt,domain=-92:0] plot ({cos(\x)}, {sin(\x)});
   \draw [black,dashed,line width=1.5pt,domain=0:89] plot ({cos(\x)}, {sin(\x)});
	\draw[latex-] [black,dashed,line width=1.5pt,domain=89:180] plot ({cos(\x)}, {sin(\x)});
	 \draw[-latex] [line width=1.5pt, black ]  (1,0) -- (1.8,0);
    \draw [line width=1.5pt, black ] (1.5,0) -- (2.5,0);
    \draw[latex-] [line width=1.5pt, black ]  (-1.8,0) -- (-2.5,0);
    \draw [line width=1.5pt, black ]  (-1,0) -- (-1.9,0);
    \node  at (0,1.3) {$(M=\pi,\sigma)$};
    \node  at (1,0.8) {$(p)$};
    \node at (2.,0.3) {$Q$};
    \node at (2.,-0.3) {$(k)$};
    \node at (0,-1.3) {$Q\;\;(p-k)$};
    \node at (-2.,0.3) {$Q$};
    \node at (-2.,-0.3) {$(k)$};
\end{tikzpicture}  
\caption{\footnotesize A schematic one--loop diagram diagram of quark--meson loops taken from \cite{Tawfik:2016ihn}.  \label{digrame} }
\end{figure}
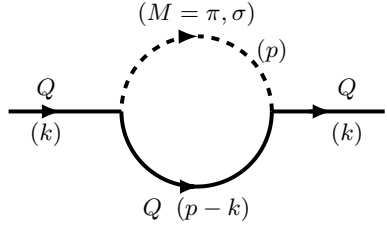
\end{center}

Figure \ref{digrame} depcits one--loop diagram of quark--meson loops. For PLSM, section \ref{sec:PLSMbv}, we limit the discussion to $\pi$ and $\sigma$ mesons, which can then be obtained from the two--point correlation function of the viscous stress--tensor, at vanishing frequency and momentum \cite{Ghosh:2014yea}. The dashed lines stands for the quark propagators, which have a finite thermal widths. The thermal width can be deduced from the quark self--energy diagram \cite{Grozin:2005yg}. In PLSM, section \ref{sec:PLSMbv}, the shear viscosity, Eq. (\ref{eq:matrix_field_A}), can be rewritten as \cite{Tawfik:2016ihn}
\bea
\eta &=& \lim_{\omega\rightarrow 0^+} \lim_{|{\bf p}|\rightarrow 0^+} \frac{\mbox{Im}\, \Pi_{11} (\omega, p)}{10\omega} = \frac{1}{10} \lim_{\omega\rightarrow 0^+} \lim_{|{\bf p}|\rightarrow 0^+} \mbox{Im}\, \Big[\sum_f  \int \frac{dk}{2\pi} \frac{(-N)}{E_{f}(k) E_{f}(p+k)}  \nonumber \\ & & \lim_{\Gamma\rightarrow 0} \left( \frac{C^-/\omega}{\left[\omega -E_{f}(k) +E_{f}(p+k)\right]+i \Gamma}+\frac{C^+/\omega}{\left[\omega +E_{f}(k) -E_{f}(p+k)\right]+i \Gamma}\right)\Big], \label{sheavs3} \\
\eta (T, \mu) &=& \frac{2}{15T}  \sum_f \int \, \frac{d^3p}{(2\pi)^3} \, \frac{|\vec{p}|^4 \tau_f}{E_f^2} \, \, f_{f} (T, \mu) \Big[1 + f_{f} (T, \mu)  \Big], \label{sheavs4}
\eea
where  $C^\mp = \mp{f_{f}(k)}^\mp  + {f_{f}(p+k)}^\mp \left[\mp\omega + E_{B,f}(k)\right]$. 

Then, the bulk viscosity, at finite temperature, $T$, and chemical potential, $\mu$, reads \cite{Tawfik:2016ihn}
\bea
\zeta (T,\mu) &=& \frac{3}{2T} \sum_f  \int \, \frac{d^3p}{(2\pi)^3} \, \frac{\tau_f(T,\mu)}{E_f^2} \left[\frac{|\vec{p}|^2}{3} - c_s^2\, E_f^2 \right]^2 \, f_{f} (T,\mu) \Big[1 + f_{f} (T,\mu)  \Big], \label{eq:zetaBUU}
\eea
where $f_{f} (T,\mu)$ was expressed in Eq. (\ref{fqaurk}) and $\tau_f(T,\mu)$ in Eq. (\ref{eq:tau1}).

\subsection{Quark condensates}
\label{sec:qrkCondns1}

\subsubsection{Vacuum quark condensates} 

The reason why we assume that the quark and gluon condensates either at vanishing or at finite temperatures considrablly contribute to the bulk viscosity among other thermodynamic quantities is the QCD color confinement \cite{Kapusta:1979fh}. Accordingly, the quark and guark condensates are associated with the dynamics of hadron wavefunctions, but not neccessarily exclusively with the vacuum \cite{Brodsky:2009zd}.

In late sixtieth of the last century, Gell--Mann, Oakes, and Renner have shown that the masses squared of the Nambu--Goldstone bosons are proportional to the masses of the light quarks \cite{GellMann:1968rz},
\bea
m_{\pi}^2 &=& A (m_u+m_d) + {\mathcal O}(m^2), \qquad\qquad A=\left|\frac{\langle0|\bar{\psi}\psi|0\rangle}{f_{\pi}^2}\right|_{m_u,m_d\rightarrow 0}, \label{eq:GMOR1}
\eea
where $|\langle|\bar{\psi}\psi|\rangle=|(\bar{u}u+\bar{d}d)|/2=8\, \pi\, f_{\pi}^3/\sqrt{3}$ \cite{Schumacher:2015wla}. Eq. (\ref{eq:GMOR1}) was originally derived up to terms linear in the quark masses. The perturbative corrections are conjectured to contribute to improving its uncertainty; less than $10\%$ \cite{Ioffe:2002ee}. As pointed out by Weinberg \cite {Weinberg:1966kf}, the chiral symmetry determines the low--energy interactions between the Nambu--Goldstone bosons, the pions, in terms of the pion decay constant, for instance, due to the limit that the velocities of the incoming pions become low or their center--of--mass energy approaches $m_{\pi}^2$. Thus, the amplitude of their elastic collision inclines to $\sim 3 m_{\pi}^2/f_{\pi}^2$, and the proportionality constant $A$ turns to be relating to the light quark condensates. Apart from the higher--order corrections, Eq. (\ref{eq:GMOR1}) obviously states that the masses of the low--lying Nambu--Goldstone bosons are given by the product of the quark condensates and masses. While the quark condensates measure the strength of spontaneous symmetry breaking, the quark masses themselves are - in turn - responsible for the chiral symmetry breaking in the QCD Lagrangian. Therefore, Eq. (\ref{eq:GMOR1}) can be rewritten as \cite{Tawfik:2005qh}
\bea
\langle0|\bar{q}q|0\rangle &\simeq &  \frac{m_{\pi}^2}{m_u+m_d}\,  f_{\pi}^2. \label{eq:GMOR2}
\eea
At physical decay constant $f_{\pi}=130.41\pm0.03~$MeV \cite{Tanabashi:2018oca}, $|\langle|\bar{q}q|\rangle=(338.144~\mathtt{MeV})^3$. 

Assuming isospin symmetry, the pion mass can be determined from the pole position in the two--point function under Fourier transform of $\langle 0|T\, A_{\mu}^i(x) A_{\nu}^k(y)|0\rangle$, where the operator $T$ denotes the time ordering assuring that the field operators are to be ordered so that their time arguments increase from right to left, for instance. Taking into account the higher orders, the pion mass can be expressed as
\bea
m_{\pi}^2 &=& m^2 \left[1 + \frac{m^2}{32 \pi^2 f_{\pi,\chi}^2} \ln\left(\frac{m}{\Lambda_3}\right)^2 + {\mathcal O}\left(m^4\right) \right], \label{eq:massHO1}
\eea 
where $f_{\pi,\chi}$ is the pion decay constant in the chiral limit, $m\equiv 2\, A\, m_{ud}$, with $m_{ud}$ being the mean mass of the two light quarks. $\Lambda_3=0.63\pm0.06~$GeV \cite{Aoki:2016frl,Leutwyler:2012} is the renormalization group invariant scale. From Eq. (\ref{eq:massHO1}), additional higher--orders can be added to the physical decay constant, so that
\bea
f_{\pi} &=& f_{\pi,\chi} \left[1 - \frac{m^2}{16 \pi^2 f_{\pi,\chi}^2} \ln\left(\frac{m}{\Lambda_4}\right)^2 + {\mathcal O}\left(m^4\right) \right], \label{eq:pdcHO1}
\eea
where $\Lambda_4=1.22\pm0.12~$GeV and the ratio of physical to chiral decay constant was estimated as $f_{\pi}/f_{\pi,\chi}=1.0719\pm0.005$ \cite{Aoki:2016frl,Leutwyler:2012}.

Similar to Eq. (\ref{eq:GMOR1}), other pseudoscalar mesons \cite{Tawfik:2005qh}, such as kaons and eta particles, have masses much higher than that of the Nambu--Goldstone bosons, the pions, can be given as
\bea
m_{K}^2 &=& B (m_{u d}+m_s) + {\mathcal O}\left(m^2\right), \qquad\qquad 
B=\left|\frac{\langle 0|\bar{\psi}\psi |0\rangle}{f_{K}^2}\right|_{m_{u d},m_s \rightarrow 0} 
 \equiv A - {\mathcal O}\left(m_s\right), \label{eq:GMOR3}\\
m_{\eta}^2 &=& C (m_{u d}+2 m_s) + {\mathcal O}\left(m^2\right), \qquad\quad\;\;
C=\left|\frac{\langle 0|\bar{\psi}\psi |0\rangle}{f_{\eta}^2}\right|_{m_{u d},m_s \rightarrow 0} 
 \equiv \frac{2}{3} \left[A - {\mathcal O}\left(m_s\right)\right], \label{eq:GMOR4}
\eea
where - up to the leading order - $m_{\eta}^2 =(4 m_{K}^2 - m_{\pi}^2)/3$  \cite{GellMann:1962xb}. In full lattice QCD \cite{McNeile:2012xh,Davies:2018hmw}, a recent determination of light and strange quark condensates from heavy--light current--current correlations suggests that  \cite{McNeile:2012xh,Davies:2018hmw}
\bea
\frac{\langle 0|\bar{q}q|0\rangle}{\langle 0|\bar{s}s|0\rangle} &=& \frac{(283\pm 2~\mathtt{MeV})^3}{(296\pm 11~\mathtt{MeV})^3}=0.956\pm 0.028.
\eea

\subsubsection{Quark condensates at finite temperatures} 

As thermal systems, such as high--energy collisions and the early Universe,  likely manifest the properties of strong interactions, the chiral symmetry, which as discussed can be measured by the quark condensates, significantly contributes with essential information to the partition function, from which the bulk viscosity could be - among many other thermal quantities -  derived. To this end, it is required that the possible thermal influences on the quark condensates can not be neglected, especially that of the lightest Nambu-Goldstone bosons, the pions. Calclulations up to two loops in chiral perturbation theory proposed that the temperature dependence of the light quark condensate and the pion decay constant, respectively, reads \cite{Gasser:1986vb}
\bea
\langle\bar{q}q\rangle(T) &=& \langle0|\bar{q}q|0\rangle\left[1+\frac{T^2}{8\, f^2_{\pi,\chi}} - \frac{T^4}{38\, f^4_{\pi,\chi}} + {\mathcal O}\left(T^6\right)\right], \\
f_{\pi}(T) &=& f_{\pi,\chi} \left[1-\frac{T^2}{8\, f^2_{\pi,\chi}} + {\mathcal O}\left(T^4\right)\right].
\eea
With this regard, we refer to Eq. (\ref{eq:sigmas}), which expresses the condensates of up ($\sigma_u$), down ($\sigma_d$), and strange quark ($\sigma_s$) in dependence on the orthogonal basis transformation from $\bar{\sigma}_0$, $\bar{\sigma}_3$, and $\bar{\sigma}_8$, respectively, as calculated in PLSM.  

The remaining quarks have masses heavier than the QCD scale, which is nearly of the order of the strange quark mass. These can not be treated as a small perturbation around the explicit symmetry limit, as down so far, when adding higher corrections. But up to the leading orders within the chiral limit and when taking into account SU($4$)$_L \times$ SU($4$)$_R$ symmetries \cite{Lenaghan:2000ey} only, the orthogonal basis transformation from $\bar{\sigma}_0$,  $\bar{\sigma}_8$, and  $\bar{\sigma}_{15}$, to $\sigma_l$, $\sigma_s$, and the charm quark flavor $\sigma_c$ can be expressed as 
\bea
\sigma_l &=& \frac{1}{\sqrt{2}} \bar{\sigma}_0 + \frac{1}{\sqrt{3}} \bar{\sigma}_8 + \frac{1}{\sqrt{6}} \sigma_{15}, \\
\sigma_s &=& \frac{1}{2} \bar{\sigma}_0 - \sqrt{\frac{2}{3}} \bar{\sigma}_8 + \frac{1}{2\sqrt{3}} \bar{\sigma}_{15}, \\
\sigma_c &=& \frac{1}{2} \bar{\sigma}_0 - \frac{\sqrt{3}}{2} \bar{\sigma}_{15}. 
\eea
As assumed, in the chiral limit, we have 
\bea
\sigma_{l_0} &=& f_{\pi \chi}, \\
\sigma_{s_0} &=& \frac{2f_{K \chi}-f_{\pi \chi}}{\sqrt{2}},\\
\sigma_{c_0} &=& \frac{2f_{D \chi} - f_{\pi \chi}}{\sqrt{2}}. 
\eea
For the sake of completeness, we emphasize that Eq. (\ref{pure:meson}) could be rewritten as
\begin{eqnarray}
U(\sigma_l, \sigma_s, \sigma_c) &=& - h_l \sigma_l - h_s \sigma_s - h_c \sigma_c  + \frac{m^2\, (\sigma^2_l+\sigma^2_s+\sigma^2_c)}{2} - \frac{c\, \sigma^2_l \sigma_s \sigma_c}{4}  
+ \frac{\lambda_1\, \sigma^2_l \sigma^2_s}{2}  \nonumber \\
&+& \frac{\lambda_1 \sigma_l^2 \sigma_c^2}{2} + \frac{\lambda_1 \sigma_s^2 \sigma_c^2}{2} + \frac{(2 \lambda_1+\lambda_2)\sigma^4_l }{8}   +\frac{( \lambda_1 +\lambda_2)\sigma^4_s }{4}+ \frac{(\lambda_1+\lambda_2)\sigma^4_c}{4}.\hspace*{8mm} \label{Upotio}
\end{eqnarray}
A comprehensive study for quark condensates at finite temperature in the hadron resonance gas model (HRGM) was reported in ref. \cite{Tawfik:2005qh}.

\subsection{Gluon condensates}
\label{sec:glnCondns1}

\subsubsection{Vacuum gluon condensates}
 
The gluon condensate was predicted by Shifman, Vainshtein, and Zakharov \cite{Shifman:1978by}. From the QCD sum rules for charmonium, an estimation for the renormalization invariant quantity at the lowest dimension was found finite, i.e., similar to $\langle 0|\bar{q}q|0\rangle \neq 0$, 
\bea
\left\langle 0\left|\frac{\alpha_s}{\pi} G_{\mu\nu} G^{\mu\nu}\right|0\right\rangle &=& 0.012~\mathtt{GeV}^4,
\eea
where $G^{\mu\nu}$ is the gluon field strength tensor indicating that the vacuum energy could be determined as \cite{Shifman:1978by}
\bea
\rho_0 =- \frac{9}{32} \left\langle0\left|\frac{\alpha_s}{\pi} G^2\right|0\right\rangle,
\eea
where $\alpha_s$ is the running coupling constant \cite{Deur:2016tte}. By analyzing the vacuum--vacuum current correlators as constrained by the charmoium production, for instance, $G_{\mu\nu} G^{\mu\nu}$ can be determined even empirically, especially when recalling Meissner effect and/or the gluon contributions to the higher Fock state light--front wavefunctions of hadrons \cite{Brodsky:2009zd}
\bea
G_{\mu\nu} G^{\mu\nu} &=& 2 \sum_i(|{\mathbf B}^i|^2-|{\mathbf E}^i|^2),
\eea
where ${\mathbf B}$ and ${\mathbf E}$ are magnetic and electric fields, respectively.

\subsubsection{Gluon condensate at finite temperature}
 
The relationship between the finite--temperature gluon condensates \cite{Miller:2004em,Miller:2004uc,Miller:2003ch,Miller:2003hh,Miller:2003ha} and the trace of the energy--momentum tensor \cite{Colangelo:2013ila} and therefrom the connection with the bulk viscosity suggests that
\bea
G^2(T) &=& G_0^2 \left[1-\left(\frac{T}{T_{\chi}}\right)^4 \right], \label{eq:g2c1}\\
G^2(T) &=& G_0^2 - \left[\rho(T)-3p(T)\right], \label{eq:g2c2}
\eea
where
\bea
G^2 &=& -\frac{\beta(g)}{2 g^3}\, G_{\mu\nu} G^{\mu\nu}. \label{eq:g2c3}
\eea
The renormalization group beta function and the running coupling constant, at finite temperature, are given as
\bea
\beta(g) &\simeq& -\frac{1}{48 \pi^2} \left(11\, N_c - 2\, n_f\right) g^3 + {\mathcal O}\left(g^5\right), \label{eq:g2c4}\\
\alpha_s(T) &\simeq& \frac{12\, \pi}{\left(11\, N_c - 2\, n_f\right) \, \ln\left(\frac{T}{\Lambda_{\mathtt{QCD}}}\right)^2}, \label{eq:g2c5}
\eea
where $\Lambda_{\mathtt{QCD}}$ is the QCD scale and $N_c$ and $n_f$ are the color and quark degrees of freedom, respectively.

\subsection{Quantum chromodynamics}
\label{sec:qcd}

\subsubsection{Polyakov linear-sigma model (quark condensates)}
\label{sec:PLSMbv}

According to BUU, section \ref{sec:buu}, and GK, section \ref{sec:GK}, the bulk viscosity, at least the thermal part, Eq. (\ref{eq:zetaBUU1}), is defined as a thermodynamic quantity. This might be related to the forcing causing or being generated from the expansion. This would mean that as the system expands or is compressed, its thermodynamic equilibrium is perturbated, on one hand. On the other hand, the bulk viscosity belongs to the processes trying to restore equilibrium. Apparently, these are irreversible. In this section, we discuss on how to deduce QCD thermodynamic quantities from the PLSM.

In the fifties of the last century, postulated for pion--nucleon interactions and chiral degrees--of--freedom, the linear--sigma model (LSM) \cite{GellMann:1960np} with a spinless scalar field $\sigma_a$ \cite{Schwinger:1957em} and the triplet pseudoscalar fields $\pi_a$  was conjectured to be based on theory of quantized fields which have been introduced by Schwinger \cite{Schwinger:1951xk,Schwinger:1953tb,Schwinger:1953zza,Schwinger:1953zz,Schwinger:1954zza,Schwinger:1954zz}. This effective model has real classical field having O($4$) vectors, $\vec{\Phi}=T_a(\vec{\sigma}_a, i\vec{\pi}_a)$ and $T_a=\lambda_a/2$ generators with Gell--Mann matrices $\lambda_a$. As a QCD--like approach, the chiral symmetry in LSM is conjectured to be broken explicitly by $3\times 3$ matrix field $H=T_a  h_a$, where $h_a$ are the external fields. Also, under $SU(2)_L\times SU(2)_R$ chiral transformation $\Phi\rightarrow L^{\dagger}\Phi R$, the spinless scalar fields $\sigma_a$ are finite. Their vacuum expectation values - in turn - break $SU(2)_L\times SU(2)_R$ down to $SU(2)_{L+R}$. As a result of the spontaneous symmetry breaking, the finite mean values of $\Phi$ fields, $\braket{\Phi}$, and of their conjugates, $\braket{\Phi^{\dag}}$ are generated with the quantum numbers of the vacuum with $U(1)_A$ anomaly \cite{Gasiorowicz:1969kn}. This leads to exact vanishing mean value of $\bar{\pi}_a$, the Nambu--Goldstone bosons, the pions, but assures finite mean value of $\bar{\sigma_a}$ corresponding to the diagonal generators $U(3)$ as $\bar{\sigma_0} \neq\, \bar{\sigma_3}\neq\, \bar{\sigma_8} \neq 0$. Also, the quarks are expected to gain masses, where $m_q=g f_{\pi}$, $g$ is the coupling and $f_{\pi}$ is the pion decay constant. It has been shown that $\sigma_a$ fields under chiral transformations exhibit a temperature behavior similar to that of the quark condensates, section \ref{sec:qrkCondns1}. Thus, $\sigma_a$ can be taken as order parameters for chiral phase transition \cite{Birse:1994cz,Roder:2003uz,Gallas:2009qp,Tawfik:2014gga,Wesp:2017tze} and accordingly for the QCD phase structure \cite{Tawfik:2014gga,Tawfik:2016gye,AbdelAalDiab:2018hrx,Tawfik:2019rdd}. Various thermodynamic quantities can be estimated, at finite density \cite{Tawfik:2016gye,Tawfik:2016ihn,Tawfik:2016edq,Tawfik:2019kaz}, finite magnetic fields \cite{Tawfik:2016lih,Tawfik:2016ihn,Tawfik:2017cdx,Tawfik:2019rdd}, and finite isospin asymmetry \cite{Tawfik:2019tkp}. 

The grand canonical partition function, $\mathcal{Z}$, sums the energies exchanged between particles and antiparticles, at finite temperatures ($T$) and/or densities ($\mu_f$), 
\begin{eqnarray}
\mathcal{Z}&=& \mathrm{Tr \,exp}[-(\hat{\mathcal{H}}-\sum_{f=u, d, s} \mu_f \hat{\mathcal{N}}_f)/T] = \int\prod_a \mathcal{D} \sigma_a \mathcal{D} \pi_a \int
\mathcal{D}\psi \mathcal{D} \bar{\psi} \mathrm{exp} \left[ \int_x
(\mathcal{L}+\sum_{f} \mu_f \bar{\psi}_f \gamma^0 \psi_f ) \label{eq:FE1}
\right],
\end{eqnarray} 
where the subscripts $f=[l,s,c,\cdots]$ refer to the quark flavors ($\mu_f$) is thus the corresponding chemical potential is the volume of the system of interest ($V$) and $\int_x\equiv i \int^{1/T}_0 dt \int_V d^3x$. It is conjectured that $\mu_f$ combines various types of chemical potentials. $\mathcal{L}$ is summed over the chiral LSM \cite{Tawfik:2014gga,Tawfik:2014uka,Tawfik:2014hwa} and the Polyakov Lagrangian \cite{Ratti:2005jh, Schaefer:2007pw,Roessner:2006xn,Fukushima:2008wg}, $\mathcal{L}=\mathcal{L}_{\chi}-\mathbf{\mathcal{U}} \left(\phi, \phi^{\ast}, T\right)$. The free energy, which can be derived as $\mathcal{F}=-T \cdot \log [\,\mathcal{Z}]/V$, plays a central role in thermodynamics.
\begin{eqnarray}
 \mathcal{F}  &=&  U(\sigma_l, \sigma_s) +\mathbf{\mathcal{U}}(\phi, \phi^*, T) + \Omega_{\bar{q}q}(T, \mu _f, B), \label{potential}
\end{eqnarray}
where, if we limit the discussion on light and strange quarks, 
\begin{itemize}
\item The purely mesonic potential part is given as
 \begin{eqnarray}
U(\sigma_l, \sigma_s) &=& - h_l \sigma_l - h_s \sigma_s + \frac{m^2}{2}\, (\sigma^2_l+\sigma^2_s) - \frac{c}{2\sqrt{2}} \sigma^2_l \sigma_s  + \frac{\lambda_1}{2} \, \sigma^2_l \sigma^2_s +\frac{(2 \lambda_1 +\lambda_2)}{8} \sigma^4_l  + \frac{(\lambda_1+\lambda_2)}{4}\sigma^4_s, \hspace*{8mm} 
\label{pure:meson}
\end{eqnarray}
with $\sigma_l$ and  $\sigma_s$ represent the finite--temperature and --density versions of the light and strange quark condensates, section \ref{sec:qrkCondns1}, as deduced from LSM. As discussed, the sigma fields show temperature-- and density--dependence similar to that of the quarks and therefore play the role as order parameters.

\item The Polyakov loop potentials \cite{Ratti:2005jh,Schaefer:2007pw,Roessner:2006xn,Fukushima:2008wg}, introduce gluonic degrees--of--freedom and dynamics of the quark--gluon interactions to the chiral LSM. The polynomial logarithmic parametrisation potential can be given as \cite{Lo:2013hla}
\bea
\frac{\mathbf{\mathcal{U}}_{\mathrm{PolyLog}}(\phi, \phi^*, T)}{T^4} &=&   \frac{-a(T)}{2} \; \phi^* \phi + b(T)\; \ln{\left[1- 6\, \phi^* \phi + 4 \,( \phi^{*3} + \phi^3) - 3 \,( \phi^* \phi)^2 \right]} \nonumber \\ &+& \frac{c(T)}{2}\, (\phi^{*3} + \phi^3) + d(T)\, ( \phi^* \phi)^2. \label{LogPloy}
\eea 
where $x(T)=[x_0 + x_1 \left(T0/T\right) + x_2 \left(T0/T\right)^2]/[1+x_3 \left(T0/T\right) + x_4 \left(T0/T\right)^2]$ and $b(T)= b_0\, \left(T0/T\right)^{b_1} [1-e^{b_2 \left(T0/T\right)^{b_3}}]$, with $x=(a,\,c,\,d)$. These coefficients $a$, $c$, and $d$ have been determined in ref. \cite{Lo:2013hla}, 

\item The quarks and antiquark potentials, at finite $T$ and $\mu_f$ \cite{Kapusta:2006pm}, read
\begin{eqnarray} 
\Omega_{\bar{q}q}(T, \mu _f)&=& -2\, T \sum_{f} \int_0^{\infty} \frac{d^3\vec{p}}{(2 \pi)^3} \; f_f (T,\mu)  \left\{ \ln \left[ 1+3\left(\phi+\phi^* e^{-\frac{E_f-\mu _f}{T}}\right)\, e^{-\frac{E_f-\mu _f}{T}}+e^{-3 \frac{E_f-\mu _f}{T}}\right] \right. \nonumber \\ 
&& \hspace*{43mm} \left. + \ln \left[ 1+3\left(\phi^*+\phi e^{-\frac{E_f+\mu _f}{T}}\right)\, e^{-\frac{E_f+\mu _f}{T}}+e^{-3 \frac{E_f+\mu _f}{T}}\right] \right\}. \hspace*{8mm} \label{PloykovPLSM}
\end{eqnarray}
When introducing Polyakov--loop corrections to the quark's degrees of freedom, then the corresponding  Fermi--Dirac distribution function is the one given in Eq. (\ref{fqaurk}). 

By using thermal expectation value of a color traced Wilson loop in the temporal direction  \cite{Polyakov:1978vu}, 
\bea
\Phi (\vec{x})=\frac{1}{N_c} \langle \mathcal{P}\left(\vec{x}\right)\rangle ,
\eea
then, the Polyakov--loop potential and that of its conjugate manifest QCD dynamics can be given as
\begin{eqnarray}
\phi = (\mathrm{Tr}_c \,\mathcal{P})/N_c, \qquad && \qquad 
\phi^* = (\mathrm{Tr}_c\,  \mathcal{P}^{\dag})/N_c, \label{phis}
\end{eqnarray}
where $\mathcal{P}$ is the Polyakov loop, which can be represented by a matrix in color space \cite{Polyakov:1978vu} 
\begin{eqnarray}
 \mathcal{P}\left(\vec{x}\right)=\mathcal{P}\mathrm{exp}\left[i\int_0^{\beta}d \tau A_4(\vec{x}, \tau)\right],\label{loop}
\end{eqnarray}
where $\beta=1/T$ stands for the inverse temperature and $A_4 = i A^0$ is the Polyakov gauge \cite{Polyakov:1978vu,Susskind:1979up}.

For sake of a greater elaboration, we recall that the Polyakov--loop matrix can be represented as a diagonal representation \cite{Fukushima:2003fw}. The Polyakov loop and the quarks are coupled. This is given by the covariant derivative $D_{\mu}=\partial_{\mu}-i A_{\mu}$, in which $A_{\mu}=\delta_{\mu 0} A_0$  is restricted to the chiral limit. As discussed, the PLSM Lagrangian is invariant under chiral flavor group, similar to the QCD Lagrangian \cite{Ratti:2005jh,Roessner:2006xn,Fukushima:2008wg}. 

\end{itemize}

In pure gauge limit, i.e., no quark flavors, we find that $\phi=\phi^{\ast}$ and each of them is taken as an order parameter for the QCD deconfinement phase transition  \cite{Ratti:2005jh,Schaefer:2007pw}. In order to take into account the thermodynamic behavior, we use a temperature--dependent potential $U(\phi, \phi^{\ast},T)$, Eq. (\ref{LogPloy}). When comparing the PLSM results with the lattice QCD simulations, $Z(3)$ center symmetry is found similar to that of the pure gauge QCD Lagrangian \cite{Ratti:2005jh,Schaefer:2007pw}. 

As discussed, the mean values of $\braket{\Phi}$ and that of $\braket{\Phi^{\dag}}$ are generated with the quantum numbers of the vacuum with $U(1)_A$ anomaly. Also, $\bar{\sigma_3}$ breaks the isospin symmetry SU($2$) \cite{Gasiorowicz:1969kn} and $h_a$, where $H=T_a\, h_a$. Accordingly, the diagonal components of the symmetry generators $h_0,\, h_3,\, h_8$ are finite leading to three finite condensates $\bar{\sigma_0},\; \bar{\sigma_3}\; \mbox{and}\; \bar{\sigma_8}$ and $m_u \neq m_d \neq m_s$. It would be convenient to convert the condensates by the orthogonal basis transformation from the original basis, $\bar{\sigma}_0$, $\bar{\sigma}_3$, and $\bar{\sigma}_8$ to pure up ($\sigma_u$), down ($\sigma_d$), and strange ($sigma_s$) quark flavor basis, respectively, 
\bea  
\begin{bmatrix}
     \sigma_u \\    \sigma_d \\  \sigma_s    
\end{bmatrix} = \frac{1}{\sqrt{3}} 
\begin{bmatrix}
      \sqrt{2} &1 & 1 \\
      \sqrt{2} &-1 & 1 \\
      1 & 0& -\sqrt{2} \\
\end{bmatrix} 
\begin{bmatrix}
    \bar{\sigma_0} \\    \bar{\sigma_3} \\  \bar{\sigma_8}   
\end{bmatrix}. \label{eq:sigmas}
\eea
Thus, the masses of $u$, $d$, and $s$ quarks can be given as, 
\bea
m_u &=& \frac{g}{2} \bar{\sigma}_u, \label{chiral_mass1}\\
m_d &=& \frac{g}{2} \bar{\sigma}_d, \label{chiral_mass2}\\
m_s &=& \frac{g}{\sqrt{2}} \bar{\sigma}_s.  \label{chiral_mass3}
\eea

When assuming global minimization of the free energy ($\mathcal{F}$), 
\begin{eqnarray}
\left.\frac{\partial \mathcal{F} }{\partial \sigma_l} = \frac{\partial
\mathcal{F}}{\partial \sigma_s}= \frac{\partial \mathcal{F} }{\partial
\phi}= \frac{\partial \mathcal{F} }{\partial \phi^{\ast}}\right|_{min} &=& 0, \label{cond1}
\end{eqnarray}
the remaining parameters $\sigma_l=\bar{\sigma_l}$, $\sigma_s=\bar{\sigma_s}$, $\phi=\bar{\phi}$ and $\phi^{\ast}=\bar{\phi^{\ast}}$ and their dependence on $T$ and $\mu$ could be determined. Having the thermodynamic free energy constructed, Eq. (\ref{eq:FE1}), and assuring global minimization, Eq. (\ref{cond1}), the different thermodynamic quantities can be estimated. Substituting the thermodynamic quantities in Eq. (\ref{eq:zetaBUU1}), the bulk viscosity can be evaluated, at finite $T$ and $\mu$.

\subsubsection{Lattice QCD simulations (thermodynamic bulk viscosity)}
\label{sec:LQCDbv}

Pioneering lattice QCD simulations for viscosity has been reported in ref. \cite{Sakai:2007cm}. This was possible through accumulating a large amount of configurations for {\it discrete} Green function, Eqs. (\ref{eq:Azeta}) and (\ref{eq:Aeta}) in Matsubara frequencies on isotropic $24^2 \times 8$ and $16^2 \times 8$ lattices. The viscous coefficients are determined as slopes of the spectral functions at vanishing Matsubara frequency. 
A recent estimation for the temperature dependence of the bulk viscosity of SU($3$) gluodynamics was conducted on $48^3 \times 16$ lattice \cite{PhysRevD.98.054515}. Both lattice results on $\zeta/s$ are depicted in Fig. \ref{fig:zetas}. 

Another milestone was set by ref. \cite{Kharzeev:2007wb,Karsch:2007jc}, in which Eq. (\ref{eq:spctlGK}) and its relation to the retarded Green function as defined by the Kramers--Kronkig relation and given in Eq. (\ref{sheavs3}) in terms of thermodynamic quantities. Furthermore, the authors of ref. \cite{Karsch:2007jc} took into consideration the fact that the bulk viscosity also measures the violation of the conformal invariance. Thus, it was pointed out that the QCD at the classical level is conformally invariant. This was the reason why quark and gluon condensates, sections \ref{sec:qrkCondns1} and \ref{sec:glnCondns1}, have also been proposed to contribute to the bulk viscosity,
\bea
\zeta &=& \frac{1}{9 \omega_0} \Big[T\, s \left(\frac{\partial \rho}{\partial p} -3\right) - 4(\rho - 3p)  \qquad\quad\;\,\mathtt{thermal}\;\;\mathtt{parts} \nonumber \\
&+&  \left(T\frac{\partial}{\partial T} -2 \right) \langle\bar{q}q\rangle(T) + g_g\, G^2(T)  \qquad\quad \mathtt{thermal\;q\;\&\;g\;condensates} \nonumber \\
&+& g_f\left(m_{\pi}^2 f_{\pi}^2 + m_{K}^2 f_{K}^2+ m_D^2 f_D^2 + \cdots\right)\Big].  \;\;\;\mathtt {vacuum\;q\;\&\;g\;condensates} \label{eq:zeta1}
\eea 
where $g_g$ and $g_f$ are the degeneracy factors for gluon and gluons, respectively. $g_g=16$ (spin polarization multiplied by $N_c^2-1$) and $N_c$ is the color degrees of freedom. $g_f=12\, n_f$ (spin polarization multiplied by parity multiplied by $N_c \times n_f$, where $n_f$ are the degrees of freedom of the quark flavors. $m_D$ and $f_D$ stand for mass and decay constant of $D$-meson, respectively. The scale parameter $\omega_0$ defines the applicability of parturbation theory. In Eq. (\ref{eq:zeta1}), $\zeta$ was obtained using the frequency limit of the spectral density at vanishing spatial momentum \cite{Karsch:2007jc,NoronhaHostler:2008ju}. 

At $T>T_c$, a combination of low--energy theorems, as detailed in sections \ref{sec:qrkCondns1} and \ref{sec:glnCondns1} and in the second line of Eq. (\ref{eq:zeta1}) with finite--temperature non-perturbative calculations, first line of Eq. (\ref{eq:zeta1}) was studied in ref. \cite{Kharzeev:2007wb} and later on introduced in ref. \cite{Karsch:2007jc}. At $T<T_c$, the same approach was utilized to obtain the upper bounds on the shear and bulk viscosity normalized to the entropy density.

\section{Results and discussion} 
\label{resulat} 

\subsection{Bulk viscosity in HRGM and PLSM}
\label{sec:hrmplsm}

To draw a picture for the temperature dependence of $\zeta$ around the hadron--quark phase transition, we confront our PLSM calculations with recent lattice QCD results \cite{Sakai:2007cm,PhysRevD.98.054515} and compare these with the HRGM estimations. From the convincing agreement above $T_c$, the integrability with the HRGM calculations, and the resulting parameterizations various conclusions can be drawn now. First, they affirm the certainty of the methodology utilized in the present work, namely, Eq. (\ref{eq:zeta1}) as deduced from BUU and KG approachs. It should be noticed that the KG approach was also utilized in the lattice QCD simulations \cite{Sakai:2007cm}, as elaborated in section \ref{sec:LQCDbv}. Second, they propose an essential extension to temperatures below $T_c$ and accordingly help in characterizing the possible impacts the hadron--quark phase transition on $\zeta$, which are apparently very significant. Last but not least, they motivate the attempt of the present work to cover a wider range of higher temperatures and then to initiate implications on heavy--ion collisions \cite{Bernhard:2016tnd} and on physics of the early Universe \cite{Tawfik:2019jsa}. 

It is worthy highlighting that the HRGM and PLSM calculations are based on Eq. (\ref{eq:zetaBUU}), which in turn is equivalent to the first line of Eq. (\ref{eq:zeta1}). The same was done in generating the lattice QCD calculations \cite{Sakai:2007cm}. The results, which are based on the entire Eq. (\ref{eq:zeta1}), shall be presented in section \ref{sec:bvlqcd}. 

\begin{figure}[htb]
\centering{
\includegraphics[width=8cm,angle=-90]{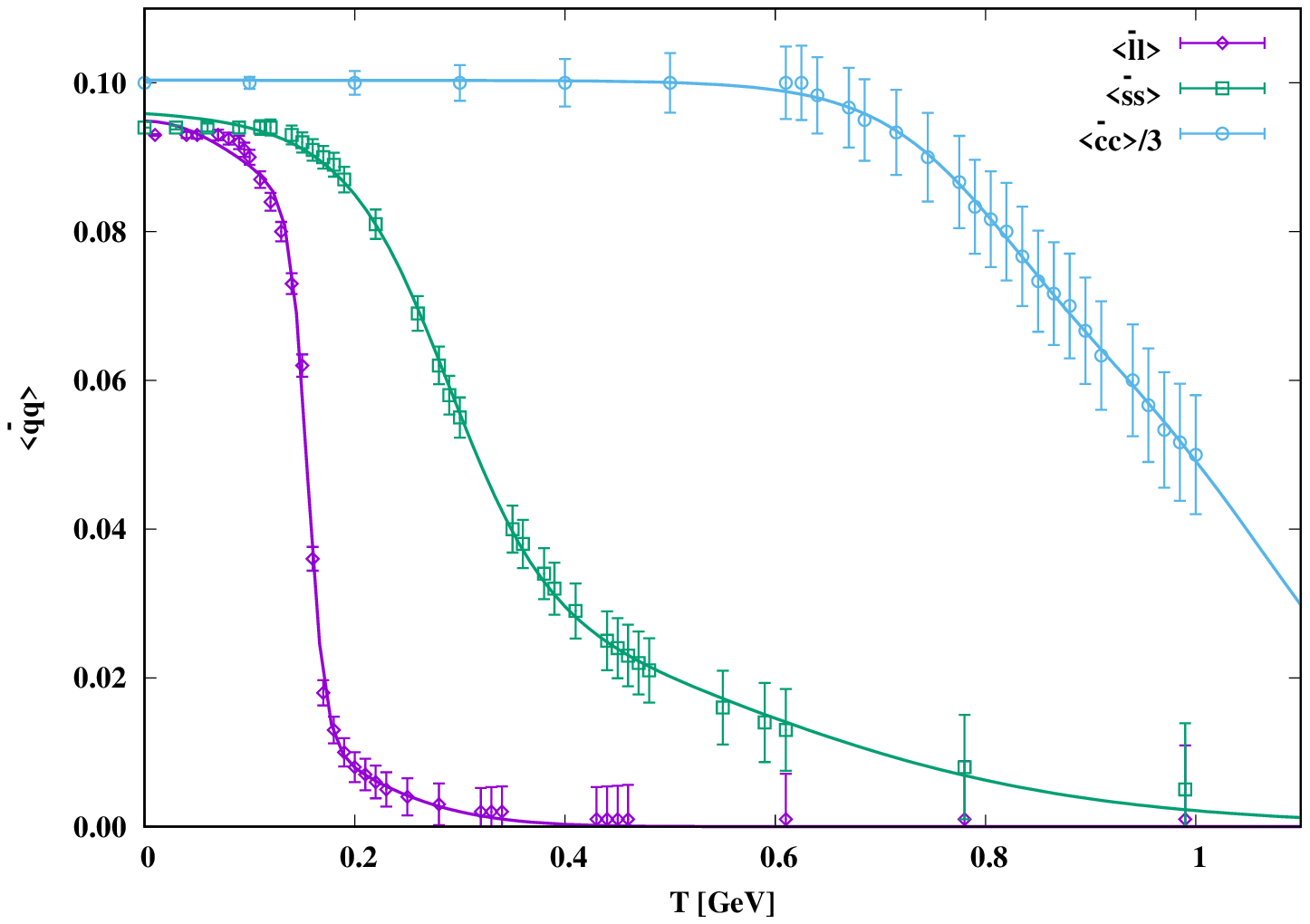}
\caption{\footnotesize The temperature dependence of light, strange, and charm quark condensates as calculated in PLSM (symbols), at vanishing baryon--chemical potential. The curves represent binomial parameterizations. 
\label{fig:qqCnds}}
}
\end{figure}

We first start with the main contributions from PLSM, namely the quark condensates. Fig. \ref{fig:qqCnds} depicts the temperature dependence of light, strange, and charm quark condensates. The symbols represent the PLSM calculations, Eq. (\ref{eq:sigmas}). The curves are binomial parametrizations. We notice that both light and strange quark condensates almost diminish at the QCD critical temperature. At high temperatures the dominant contribution to the bulk viscosity is stemming from the charm condensate. The contributions that the quarks and gluons come up with to the bulk viscosity are illustrated in top panel of Fig. \ref{fig:zeta}. Accordingly, it is likely that the still missing bottom and top quark condensates, as shall be elaborated in section \ref{sec:bvlqcd}, become dominant at higher temperatures or larger energy densities. We also notice that the temperature dependence is very structured, i.e., non--monotonic.

\begin{figure}[htb]
\centering{
\includegraphics[width=8cm,angle=-90]{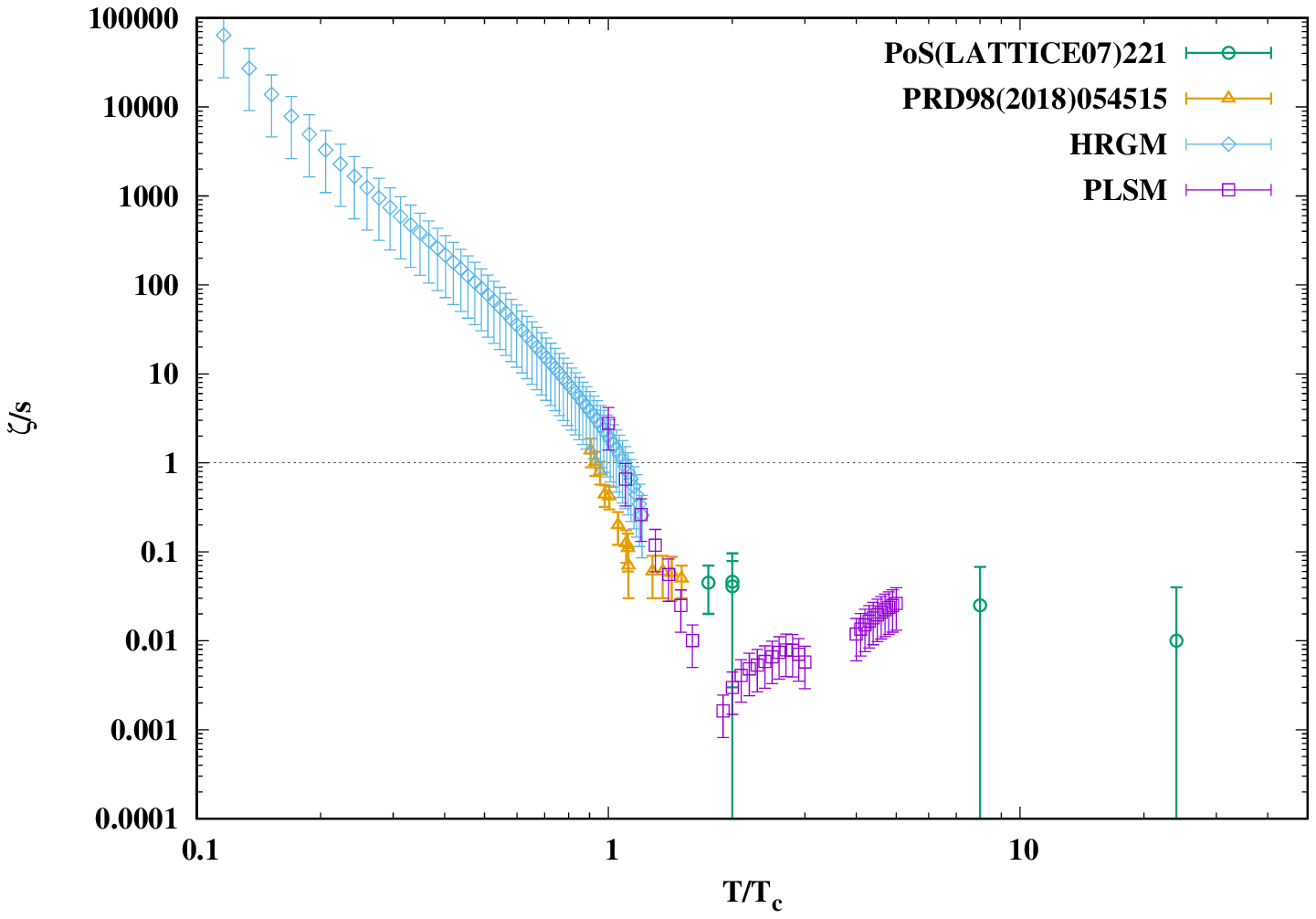}
\caption{\footnotesize At vanishing baryon--chemical potential, the temperature dependence of $\zeta/s$ as calculated in lattice QCD (circles and triangles) and PLSM (squares) above $T_c$ with the HRGM (diamants) below $T_c$. The curves present the corresponding parameterizations.
\label{fig:zetas}}
}
\end{figure}

In Fig. \ref{fig:zetas}, the dimensionless $\zeta/s$ is depicted in dependence on $T/T_c$. The results of PLSM (squares) and HRGM (diamants) are compared with the lattice QCD simulations (circles) \cite{Sakai:2007cm} and triangles \cite{PhysRevD.98.054515}. Presenting results on $\zeta/s$, which are also calculated with the same approach as the lattice QCD \cite{Sakai:2007cm}, affirms the correctness of PLSM (squares) and HRGM (diamants). 

Because of the huge decrease with increasing $T$, we draw the results in log--log scale. At $T<T_c$, the HRGM results match well with PLSM and lattice QCD. A similar result shall be reposted in section \ref{sec:bvlqcd}, in which recent lattice QCD simulations with $2+1+1+1$ quark flavors at a wide range of temperatures are taken into considration. The lattice QCD simulations \cite{Sakai:2007cm} are limited to temperatures $>T_c$. With the present study we cover $\leq T_c$ by HRGM (diamants) \cite{Tawfik:2010mb}, as well. At $T\leq T_c$, we notice that $\zeta/s$ rapidly decreases with the increase in $T$. This might be understood - among others - due to quark condensates and interaction measure, which set on their decrease and increase, respectively, with increasing temperature, especially within the region of the QCD phae transition (crossover). At $T>T_c$, the lattice QCD \cite{Sakai:2007cm} and the PLSM calculations (squares) both indicate that $\zeta/s$ drops and keeps its small value over a wide range of temperatures. The reason would be the large entropy density ($s/T^3$) and the decreasing interaction measure [$(\rho-3p)/T^4$] above $T_c$, i.e., approaching an asymptotic limit, the Stefan--Boltzmann limit.

To draw a picture about the entropy density, we recall the Stefan--Boltzmann (SB) approach. The various thermodynamic quantities can be derived from the partition function characterizing an ideal gas of free quarks and gluons \cite{Letessier:2002gp}. For example, the entropy density reads
\bea
\frac{s_{\mathtt{SB}}}{T^3} &=& \frac{2\, \pi^2}{45} g_g - \frac{1}{6} \sum_f g_f \left[\left(\frac{\mu_f}{T}\right)^2 + \frac{1}{\pi^2} \left(\frac{\mu_f}{T}\right)^4\right] +
\frac{2}{45} \sum_f g_f \left[ \frac{7}{4}\pi^2 + \frac{15}{2} \left(\frac{\mu_f}{T}\right)^2 +  \frac{15}{4 \pi^2} \left(\frac{\mu_f}{T}\right)^4\right]. \label{SBlimitsEQ} 
\eea
At vanishing $\mu_f$, $n_f=3$, $g_f=36$, and $g_f=16$, $s_{\mathtt{SB}}/T^3\simeq 34.62$ while $(\rho-3p)/T^4\rightarrow 0$. Therefore, at high temperatures, $\zeta/s$ tends to very small values. 

In light of this, the physical meaning of the $\zeta/s$ reflects with it the impacts of the entropy. Thus, $\zeta/s$ could be interpreted as bulk viscosity per degree of freedom. The latter varies from phase (hadron) to phase (QGP). In the section that follows, we focus of $\zeta$, at varying energy density.

\subsection{Bulk viscosity in non-perturbative and perturbative calculations}
\label{sec:bvlqcd}

\subsubsection{QCD contributions}
\label{sec:QCDconts}

\begin{figure}[htb]
\centering{
\includegraphics[width=8cm,angle=-90]{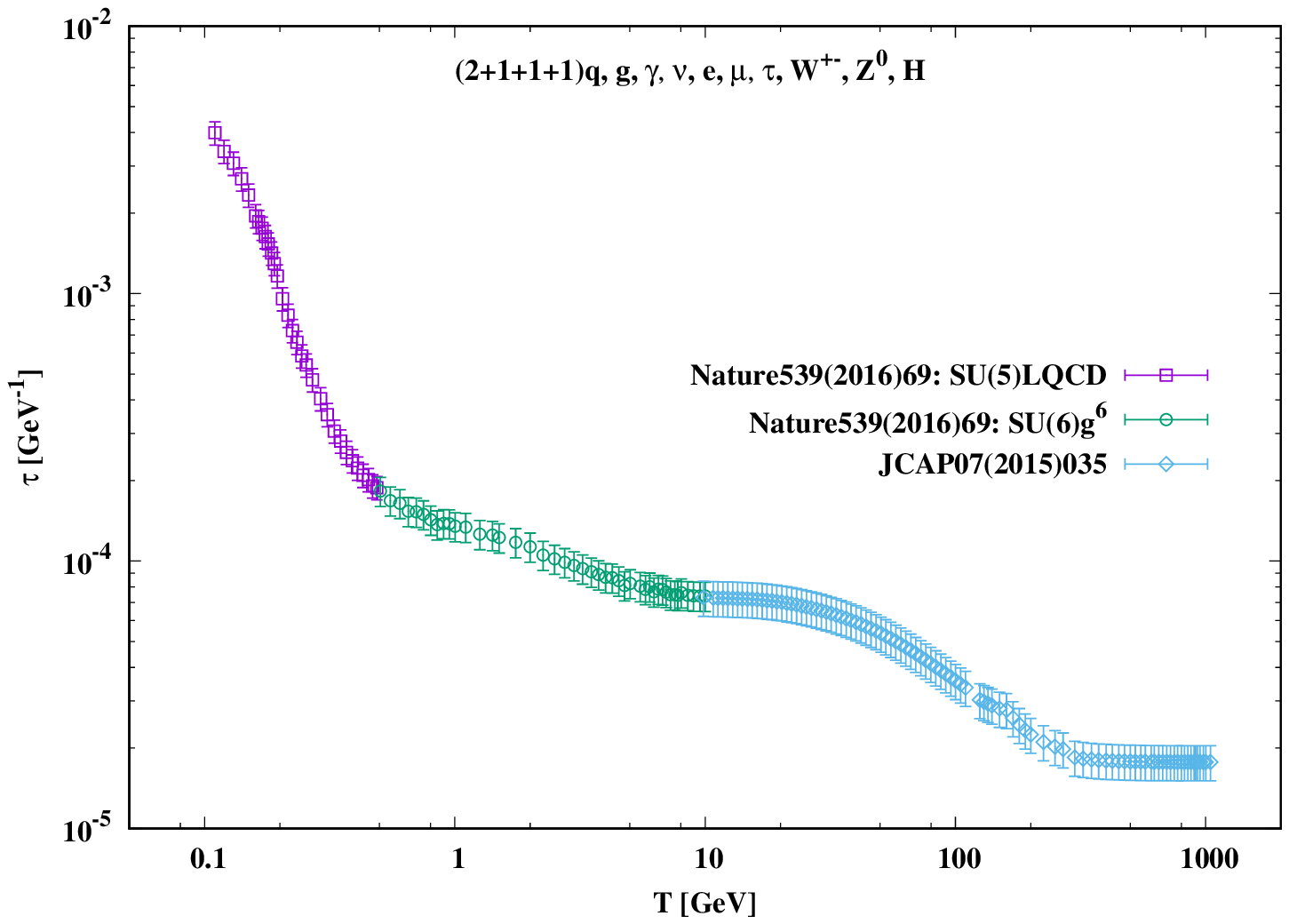}
\caption{\footnotesize The temperature dependence of $\tau$ is presented. The results are calculated, at vanishing baryon--chemical potential, and estimated from non--perturbative and perturbative QCD simulations (bottom symbols), to which quark and gluon condensates and thermodynamics of gauge bosons, charged leptons, and Higgs bosons are added (left symbols).
\label{fig:tau}}
}
\end{figure}

The relaxation time $\tau_f(T)$ plays an essential role in estimating $\zeta$. As introduced in section \ref{sec:buu}, $\tau_f(T)$ involves complicated collision integrals. Using cross section and mean collision time, i.e., thermal averages \cite{Tawfik:2016edq,Tawfik:2011sh,Tawfik:2010bm} led to Eq. (\ref{eq:tau1}). Fig. \ref{fig:tau} depicts the temperature dependence of $\tau$ as calculated in non--perturbative and perturbative QCD simulations \cite{Borsanyi:2016ksw,Laine:2015kra,DOnofrio:2015gop,Tawfik:2019jsa} (bottom symbols) to which contributions from quark and gluon condensates, section \ref{sec:qrkCondns1} and \ref{sec:glnCondns1}, respectively, and thermodynamics from gauge bosons, charged leptons, and Higgs bosons are added (left symbols). We notice that $\tau$ steadily decreases with increasing $T$.  In the different phases, there are different rates of decreasing $\tau$. This figure illustrates the essential contributions of the present manuscript. 

\begin{figure}[htb]
\centering{
  \includegraphics[width=8cm,angle=-90]{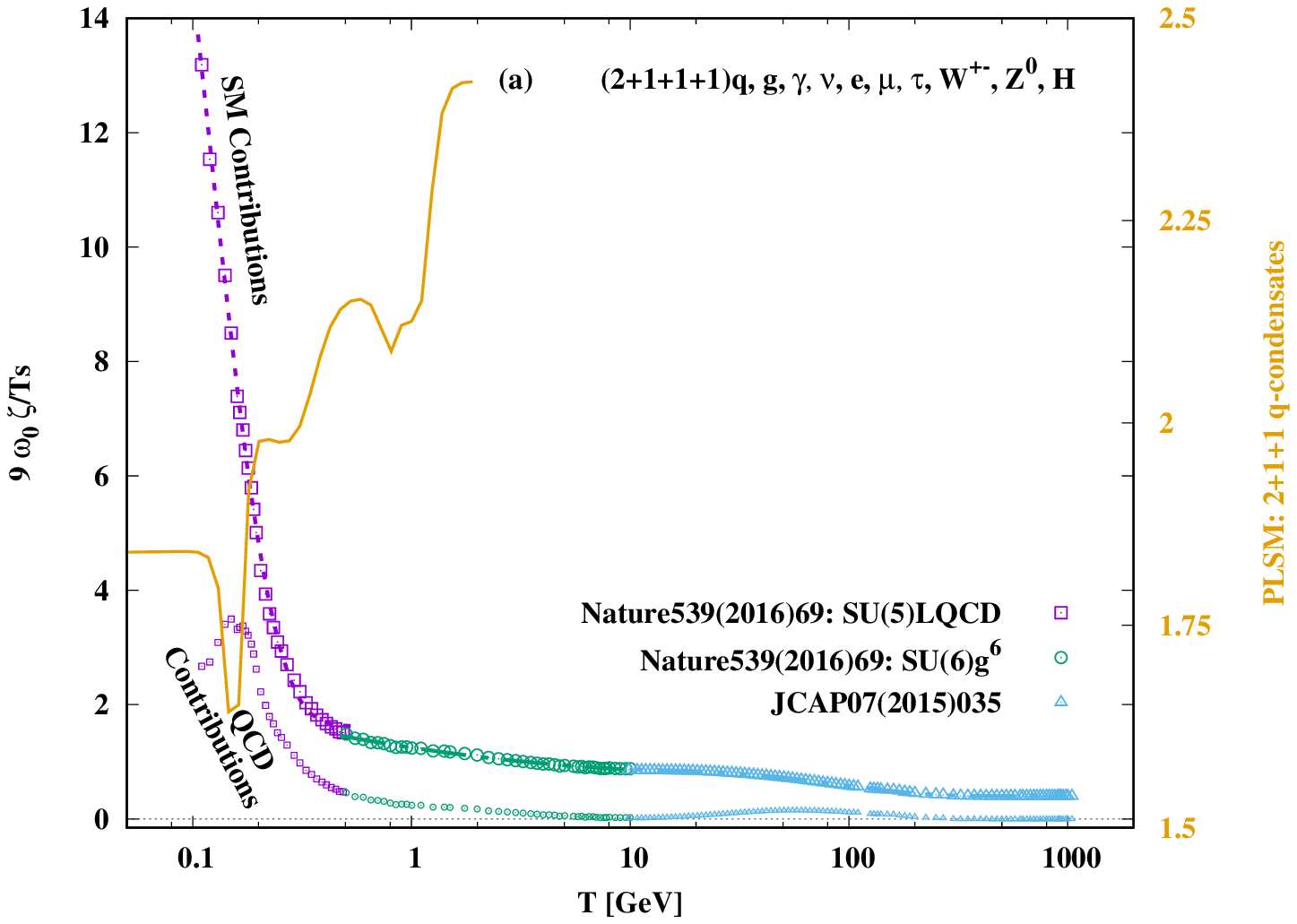}\\
  \includegraphics[width=8cm,angle=-90]{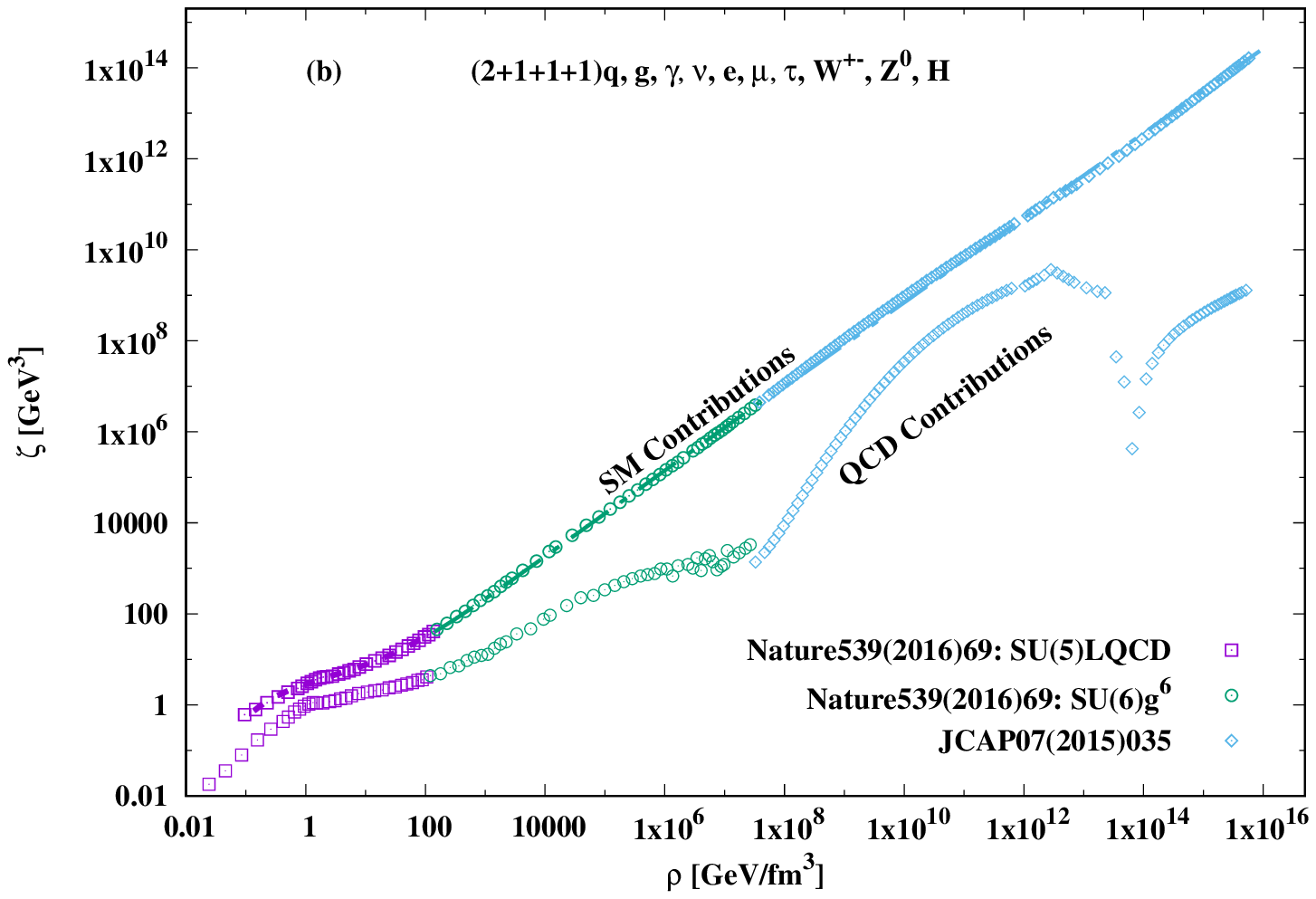}
\caption{\footnotesize Top panel depicts the dimensionless $9 \omega_0\zeta/Ts$ vs. $T$ in GeV units. Bottom panel shows the bulk viscosity $\zeta$ in dependence of the energy density $\rho$. Both quantities are calculated, at vanishing baryon--chemical potential and given in physical units. The top symbols stand for the SM contributions, section \ref{sec:SMconts}, while the bottom ones stand for the QCD contributions. The solid curve (top panel) represent the vacuum and thermal gluon and quark ($u$, $d$, $s$, and $c$) condensates, sections \ref{sec:qrkCondns1}, \ref{sec:glnCondns1}, and Fig. \ref{fig:qqCnds}. }.
\label{fig:zeta}}
\end{figure}

Top panel of Fig. \ref{fig:zeta} shows the temperature dependence of $9\omega_0\zeta/Ts$. Here, we cover temperatures ranging from $100~$MeV up to $1~$TeV. The bulk viscosity is calculated according to Eq. (\ref{eq:zeta1}), in which the thermodynamic quantities, $s/T^3$, $c_s^2$ and $(\rho-3p)/T^4$ are taken from non--perturbative \cite{Borsanyi:2016ksw} and perturbative QCD calculations \cite{Laine:2015kra,DOnofrio:2015gop}. As elaborated in ref. \cite{Tawfik:2019jsa}, both sets of calculations are properly rescalled. It was assumed that the non--perturbative effects \cite{Borsanyi:2016ksw} characterize the resulting thermodynamics, at temperatures $\leq 10~$GeV, especially that the heavier quarks have been also included. At higher temperatures, the thermodynamic quantities such as pressure, energy density, and entropy density can also be calculated, perturbatively. The results reported in refs. \cite{Laine:2015kra,DOnofrio:2015gop} cover temperatures up to $1~$TeV. With a scaling proposed in ref. \cite{Tawfik:2019jsa}, both calculations become matching with each others, perfectly smoothly. These allow the temperatures to go over the TeV-scale. It was found that the analysis using the lattice simulations \cite{Borsanyi:2016ksw,Laine:2015kra,DOnofrio:2015gop} consistently characterizes strong and EW domains. Accordingly, both have crossover transitions. From the phenomenological point of view, it was highlighted that the QCD phase transition (strongly--interacting matter) seems stronger than the thermal EW phase transition (electroweakly--interacting matter). With the regard of possible implications, the temperature dependence of $9\omega_0 \zeta/Ts$ can be applied on heavy--ion collisions \cite{Bernhard:2016tnd}, for instance. 

The reason that we start with $9\omega_0 \zeta/Ts$ vs. $T$ in GeV units, is our intention to compare with refs. \cite{Karsch:2007jc,NoronhaHostler:2008ju}. Indeed, above $T_c$, the quantity $9\omega_0 \zeta/Ts$ rapidly declines. In the present script, we go beyond the limit of $3-4\; T_c$ as done in ref. \cite{Karsch:2007jc}. We use $2+1+1+1$ lattice QCD and add our estimation for the temperature dependence of up--, down--, strange--, and charm--quark condensates from PLSM, Fig. \ref{fig:qqCnds}, together with the gluon condensates, Eqs. (\ref{eq:g2c1})-(\ref{eq:g2c5}). The temperature dependence of the quark and gluon condensates are also depicted as the solid curve in the top panel of Fig. \ref{fig:zeta}, whose y--axis is positioned to the right. This figure presents two quantities, the one in right y--axis; $9 \omega_0\zeta/T s$, while the left y--axis the vacuum and thermal gluon and quark (u, d, s, and c) condensations. Despite the limitation up to the charm quark, it is obvious that the contributions from the quark and gluon condensates are responsible for the relative large $9\omega_0 \zeta/Ts$ comparing to the results reported in ref. \cite{Karsch:2007jc}. Up to $\sim 2~$GeV, the temperature dependence of the four quark condensates vanishes. At temperatures larger than $\sim 2~$GeV, condensates of heavier quarks likely become dominant. Also, in this limit, the temperature dependence of the gluon condensates vanishes, as the interaction measure becomes very small, Eq. (\ref{eq:g2c2}), at least within the QCD sector, i.e., the system approaches the Stefan-Boltzmann limit.

Similar to the peak at the QCD crossover, there is a signature for EW crossover at about $60~$GeV. As pointed out in Fig. 1 of ref. \cite{Tawfik:2019jsa}, the EW crossover, as the name says, seems to take place within a wide range of temperatures, from $\sim 20$ to $\sim 100~$GeV. That $9\omega_0 \zeta/Ts$ is conjectured to play the role of an order parameter is comprehend, as it strongly depends on the thermodynamic quantities, entropy, speed of sound squared, and the interaction measure besides the quark and gluon condensates. Each of them reflects rapid change when going through phase transition. The reason why $9\omega_0 \zeta/Ts$ nearly vanishes below $\sim 20$ and above $\sim 100~$GeV would the absence of bottom and top quark condensates. They likely heighten the values of $9\omega_0 \zeta/Ts$ including the peak at about $60~$GeV, as well.

The bottom panel shows the bulk viscosity $\zeta$ as a function of the energy density $\rho$. Both quantities are given in physical units. Such a barotropic dependence can be straightforwardly applied in various cosmological aspects \cite{Tawfik:2011sh,Tawfik:2010pm,Tawfik:2010bm,Tawfik:2009mk} in the way that $\rho$ can be directly substituted by the Hubble parameter. We compare between QCD (bottom curve) and SM contributions (left curve). The earlier counts for the thermodynamic quantities calculated in lattice QCD \cite{Borsanyi:2016ksw,Laine:2015kra,DOnofrio:2015gop,Tawfik:2019jsa}. The latter takes into account contributions from quark and gluon condensates and thermodynamics of an ideal gas of gamma, charged leptons, $W^{\pm}$, $Z^0$ and $H$ bosons. 

We notice that different than the temperature dependence (top panel), here the dependence of QCD $\zeta$ on $\rho$ is very structured, i.e., a non--monotonic dependence. At least, there are four domains to be distinguished. The first one is the hadron--QGP phase (Hadron--QGP). This region spans over $\rho\lessapprox 100~$GeV/fm$^3$. In the second phase, $\zeta$ reaches another maximum. Here, $\rho$ covers up to $\sim 5\times 10^7~$GeV/fm$^3$. Accordingly, it would be assumed that this domain combines QCD and EW phases (QCD). The third phase seems to form an asymmetric parabola (EW), where the focus is likely positioned at the corresponding critical energy density, $\rho_c\simeq 10^{12}~$GeV/fm$^3$. The fourth region shows a rapid increase in $\rho$ emerging from non--continuous point. It seems very likely that asymmetric parabola can be constructed in each region.  

First, Hadron-QGP is characterized by a rapid increase in $\zeta$, i.e., $\zeta\eqsim 1~$GeV$^3$, at $\rho\simeq 1~$GeV/fm$^3$. This is then followed by a slight increase in $\zeta$. For example, at $\rho\simeq 100~$GeV/fm$^3$, $zeta$ becomes to $\sim130~$GeV$^3$. One could estimate that the hadron phase would be defined by $\rho\lessapprox 0.5~$GeV/fm$^3$ \cite{Tawfik:2004sw,Tawfik:2004vv}, at which $\zeta\lessapprox 0.5~$GeV$^3$. On the other hand, QGP is accommodated, at $0.5\lessapprox \rho \lessapprox 100~$GeV/fm$^3$, i.e., much wider $\rho$ than the hadron phase. An essential conclusion could be drawn now. Over this wide range of $\rho$, the bulk viscosity is obviously not only finite but large, which could be alligned with the RHIC discovery of strongly correlated QGP \cite{Ryu:2017qzn,Heinz:2011kt,Gyulassy:2004zy}. 

At higher energy densities, there is a tendency of an increase in $\zeta$ with increasing $\rho$. A peak is nearly positioned at $\rho\simeq 5\times 10^{6}~$GeV/fm$^3$ and $\zeta\simeq 200~$GeV$^3$. This would refer to a smooth transition from QCD to EW matter, i.e., from strong to electroweak matter. Precise estimation for the critical quantities and suitable order parameters shall be subject of a future study. To summarize, we recall that over a region, where $\rho$ gets an huge increase of about five--order--of--magnitude; $10^2\lessapprox\rho\lessapprox 5\times 10^{7}~$GeV/fm$^3$, there a small increase followed by a small decrease in $\zeta$ to be quantified. 

The huge jump in $\zeta$ takes place when $\rho$ increases from nearly $10^8$ to approximately $10^{12}~$GeV/fm$^3$. The corresponding bulk viscosity increases from nearly $3\times10^2$ to approximately $5\times10^9~$GeV$^3$. A further increase in $\rho$ of about one--order--of--magnitude seems not affecting $\zeta$. But with the continuation of the increasing of $\rho$ of about two--order--of--magnitude, we notice that $\zeta$ declines to $\sim 3\times 10^{5}~$GeV/fm$^3$.

\subsubsection{SM contributions}
\label{sec:SMconts}

Besides gluons and ($2+1+1+1$) quarks, the contributions of the gauge bosons; the photons, $W^{\pm}$, and $Z^0$, the charged leptons; neutrino, electron, muon, and tau, and the Higgs bosons; scalar Higgs particle, are also taken into account \cite{Tawfik:2019jsa}. Obviously, as much as possible SM contributions are included in the present calculations. The vacuum and thermal bottom quark condensate, the entire gravitational sector, neutral leptons, and top quark are the missing SM--contributions. The results are shown in Fig. \ref{fig:zeta}, as well. An overall conclusion can be drawn now. The SM contributions are very significant over the entire ranges of temperatures and energy densities. On one hand, they allow to cover higher temperatures and larger energy densities. On the other hand, the characteristic structures observed with the QCD contributions, section {sec:QCDconts}, is almost removed, so that the dependence on temperature (top panel) and on energy density (bottom panel) becomes almost monotonic. Last but not least, the temperature dependence of $9 \omega_0 \zeta/Ts$ is exponentially decreasing, while the energy--density dependence of $\zeta$ is almost linearly increasing.

For $9 \omega_0 \zeta/Ts$, the resulting parameterizations are
\bea
\mathtt{Hadron-QGP:} && \frac{9 \omega_0 \zeta}{Ts}= a_1+a_2\, \exp\left[-a_3\left(T^{a_4}\right) \right], \label{eq:qcdpt}\\
\mathtt{QCD:}   && \frac{9 \omega_0 \zeta}{Ts} = b_1+b_2\, \exp\left[-b_3\left(T^{b_4}\right) \right], \label{eq:qcdew}\\
\mathtt{EW:}   && \frac{9 \omega_0 \zeta}{Ts} = c_1 +c_2 T + c_3 \exp\left[-c_4\left(T^{c_5}\right) \right]. \label{eq:eqpt}
\eea
For Hadron--QGP: $a_1=1.624 \pm 0.054~$GeV, $a_2=29.004\pm 2.94~$GeV, $a_3=22.776\pm 2.829$, and $a_4=1.451\pm 0.101$. For QCD: 
$b_1=0.7629\pm 0.081~$GeV, $b_2=4.811\pm 1.613~$GeV, $b_3=2.291\pm1.432$, and $b_4=0.226\pm 0.174$. 
For EW: $c_1=0.404\pm 0.003~$GeV, $c_2=-5.827\times 10^{-6} \pm 1.611\times 10^{-6}$,  $c_3=0.490\pm0.005~$GeV, $c_4=0.0036\pm 0.0003$, and $c_5=1.218   \pm 0.021$. There is a rapid decrease in $9 \omega_0 \zeta/Ts$, at temperatures characterizing the hadronic matter, Eq. (\ref{eq:qcdpt}). With a slower rate, this seems to continue within the QGP, at temperatures up to $sim10~$GeV, Eq. (\ref{eq:qcdew}). At $T>10~$GeV, the decrease combines linear and exponential functions, Eq. (\ref{eq:eqpt}).

For $\zeta(\rho)$, we distinguish three regions with the parameterizations 
\bea
\mathtt{Hadron--QGP:} &&  \zeta= d_1+d_2 \rho+d_3 \rho^{d_4}, \label{eq:qcdpt2} \\
\mathtt{QCD:}   && \zeta= e_1 +e_2 \rho^{e_3}, \label{eq:qcdew2}\\
\mathtt{EW:}  && \zeta= f_1 +f_2 \rho^{f_3}. \label{eq:eqpt2}
\eea
For Hadron--QCD: $d_1= -9.336\pm 4.152$, $d_2=0.232\pm 0.003$, $d_3=11.962\pm 4.172$, and $d_4=0.087\pm 0.029$.
For QCD: $e_1= 8.042\pm 0.056$, $e_2= 0.301\pm 0.002$, and $e_3= 0.945\pm 0.0001$.
For EW: $f_1= 0.350\pm 0.065$,  $f_2= 10.019 \pm 0.934$, and $f_3= 0.929 \pm 8.898\times 10^{-5}$.

\section{Conclusions} 
\label{conclusion} 

Comparing our HRGM-- and PLSM--results of bulk viscosity with the first--principle QCD calculations, we conclude a convincing agreement, at temperatures exceeding the QCD scale and an excellent integrability below this range of temperatures. Hence, the methodology utilized in the present work, namely BUU and KG approaches, are confirmed. 

Allowing the temperature to increase from a few MeV up to TeV and the energy density up to $10^{16}~$GeV/fm$^3$, almost all possible contributions to the bulk viscosity are taken into consideration. The first type of contributions represents thermodynamic quantities calculated in non--perturbation and perturbation QCD with up, down, strange, charm, and bottom quark flavors and, of course, the entire gluonic sector. Taking into account contributions of the gauge bosons; photons, $W^{\pm}$, and $Z^0$, the charged leptons; neutrino, electron, muon, and tau, and the Higgs bosons; scalar Higgs particle, shows that these Standard Model particles are very significant, especially when bearing in mind cosmological implications, for instance. When comparing the results of bulk viscosity with and without this sector, we conclude that both temperature and energy--density dependences become to a great extend monotonic. Furthermore, this sector considerably adds to the results so that the energy density approaches about two--order--of--magnitude GeV/fm$^3$ larger than the perturpation theory. The vacuum and thermal condensations of gluons and quarks (up, down, strange, and charm) as calculated in the Polyakov linear--sigma model are the third type of contributions.  

We conclude that the bulk viscosity increases almost linearly with increasing energy density. Opposite to such a dependence is the one of the dimensionless quantity $9 \omega_0 \zeta/Ts$ on temperature, where $\omega_0$ is a perturbative scale and $s$ is the entropy density. Here, an almost linearly decrease with increasing temperature was obtained.

\section*{Acknowledgements}
The work of AT was supported by the ExtreMe Matter Institute (EMMI) at the GSI Helmholtz Centre for Heavy Ion Research.


\bibliographystyle{aip}

\bibliography{2019_10_10_Bulk_Viscosity_LQCD_PLSM} 

\end{document}